\documentclass[twocolumn]{aastex7}
\usepackage{apjfonts}
\usepackage{amsmath}
\usepackage{enumitem} 

\shorttitle{Unbiased Latent-Inclination Inference of the Tully-Fisher Relation}
\shortauthors{Fu}
\journalinfo{ApJ accepted 2025 August 7}

\newcommand{\kms}{{km s$^{-1}$}}
\newcommand{\HI}{H\,{\sc i}}
\newcommand{\OII}{[O\,{\sc ii}]}
\newcommand{\hunit}{{km\,s$^{-1}$\,Mpc$^{-1}$}}

\begin{document}
\title{Mitigating Eddington and Malmquist Biases in Latent-Inclination Inference of the Tully-Fisher Relation}
\author[0000-0001-9608-6395]{Hai Fu}
\affil{Department of Physics \& Astronomy, University of Iowa, Iowa City, IA 52242, USA}
\email{hai-fu@uiowa.edu}

\begin{abstract}
The Tully-Fisher relation is a vital distance indicator, but its precise inference is challenged by selection bias, statistical bias, and uncertain inclination corrections. This study presents a Bayesian framework that simultaneously addresses these issues. To eliminate the need for individual inclination corrections, inclination is treated as a latent variable with a known probability distribution. To correct for the distance-dependent Malmquist bias arising from sample selection, the model incorporates Gaussian scatter in the dependent variable, the distribution of the independent variable, and the observational selection function into the data likelihood. To mitigate the statistical bias---termed the ``general Eddington bias''---caused by Gaussian scatter and the non-uniform distribution of the independent variable, two methods are introduced: (1) analytical bias corrections applied to the dependent variable before likelihood computation, and (2) a dual-scatter model that accounts for Gaussian scatter in the independent variable within the likelihood function. The effectiveness of these methods is demonstrated using simulated datasets. By rigorously addressing selection and statistical biases in a latent-variable regression analysis, this work provides a robust approach for unbiased distance estimates from standardizable candles, which is critical for improving the accuracy of Hubble constant determinations.
\end{abstract}


\section{Introduction} \label{sec:intro}

The Tully-Fisher relation (TFR) is an empirical correlation between luminosity (or mass) and maximum rotation velocity of spiral galaxies \citep{Tully77}. The tightness of the correlation makes it a remarkable distance indicator in the category of standardizable candles, so it was immediately utilized to measure the Hubble constant $H_0$ \citep{Sandage76, Tully77}. In fact, the discovery of the correlation between peak luminosity and rate of decline for Type Ia supernovae, the most popular standardizable candles today, was made possible by distances from the TFR and surface brightness fluctuations \citep{Phillips93}. But it was soon realized that the TF distances were affected by local peculiar velocities \citep{Tully88} and the Malmquist bias \citep{Giraud87}. The classic Malmquist bias \citep{Eddington14, Malmquist22} is caused by observational selection effects and the intrinsic scatter in the luminosity of standard candles. The intrinsic scatter of the TFR is expected, as the physical properties of the galaxies are unlikely to be fully captured by rotation velocity alone. When a sample is subdivided in redshift, the Malmquist bias starts to show distance-dependent behaviors, causing the artifact that the mean luminosity at fixed rotation velocity (and $H_0$ as a consequence) increases outward \citep{Sandage94a, Sandage94b}.

Multiple methods have been proposed to correct for the distance-dependent Malmquist bias: (1) the method of replacing the linewidth-predicted mean luminosity with the analytical mean luminosity that depends on both flux limit and redshift \citep{Sandage94a, Butkevich05}, (2) the method of normalized distances, which enhances the S/N of the luminosity-redshift diagram by shifting galaxies of different luminosities diagonally along the luminosity limit and uses the plateau at lower normalized distances to obtain an unbiased luminosity estimate \citep{Bottinelli86, Bottinelli88}, and (3) the method of maximum likelihood estimation (MLE) that includes the distance-dependent sample selection function in the data likelihood function \citep{Willick94, Willick97}. Among the three, the MLE is the most powerful and versatile method, because it utilizes the full data set, naturally accounts for heteroscedastic measurement errors, and fits the slope, the intercept, and the intrinsic scatter simultaneously. \citet{Willick97} derived the likelihood functions for the two commonly used ``unidirectional'' models\footnote{These unidirectional models include the error and the intrinsic scatter of the dependent variable but neglect the scatter of the independent variable.}: the forward TFR and the inverse TFR, where the independent variable is velocity width and luminosity, respectively. Recent applications of the MLE method on the Cosmicflows-4 (CF4) data include \citet{Kourkchi22} for the inverse TFR and \citet{Boubel24a} for the forward TFR. However, even with identical data set, the inferred TF parameters from the two unidirectional models disagree well beyond the inferred statistical uncertainties. This happens because there is a separate bias in linear regression that stems from the independent variable.

It has long been recognized that linear regression coefficients are biased when the independent variable is measured with error, but the model ignores this error \citep[e.g.,][]{Fuller87}. In \S\,\ref{sec:bias}, I will derive the analytical formula of this additional bias, which originates from the scatter and the distribution function of the independent variable. Because the result resembles that of the classic Eddington bias of the mean true value, it will be referred to as the ``general Eddington bias''. The popular methods to mitigate this bias in general linear regression include (1) the bivariate correlated errors and intrinsic scatter (BCES) estimator \citep{Akritas96} and (2) the Bayesian dual-scatter methods (LINMIX\footnote{\url{https://github.com/jmeyers314/linmix}} by \citealt{Kelly07} and ROXY\footnote{\url{https://github.com/DeaglanBartlett/roxy}} by \citealt{Bartlett23})\footnote{While LINMIX requires the user to set (1) the Gaussian scatter of the independent variable and (2) a reasonable number of Gaussians to model the distribution function, ROXY can infer these parameters from the data.}.
The former uses the moments of the data to estimate regression coefficients by correcting the bias in the ordinary least squares (OLS) estimator, while the latter derives the data likelihood function that includes not only the intrinsic scatter and selection function of the dependent variable, but also the covariance matrixes of measurement errors and the intrinsic distribution of the independent variable (approximated there as a mixture of Gaussians for flexibility and mathematical convenience). 
The dual-scatter methods can reduce the general Eddington bias but cannot address the distance-dependent Malmquist bias, as these linear regression methods do not treat distance as a separate variable within the dependent variable. For example, in \citet{Kelly07} the sample selection function $p(I=1 | x,y)$ depends only on $x$ and $y$; while for a flux-limited survey, the selection function of $y$ should depend on distance (which is part of $y$ if it represents luminosity). 

Besides the distance-dependent Malmquist bias and the general Eddington bias, a major data-related issue limits the precision of our statistical inference of the TFR. Before modeling, the observational data must be corrected for galaxy inclination. Inclination correction is necessary because (1) the observed velocity width is projected along the line of sight and the observed luminosity is attenuated by dust in the plane of the disk and (2) random inclinations cause non-Gaussian scatters in both axes. But inclination estimation with axial ratio is highly uncertain even for well resolved galaxies, because (1) the axial ratio depends the method (isophote fitting vs. surface brightness modeling), the depth of the photometric data, and the bulge-to-disk ratio of the galaxy, and (2) the inclination angle also depends on the unknown edge-on thickness of the disk. As a result, galaxies with lower inclinations (e.g., $<45^\circ-60^\circ$ from face-on) are often discarded in TF studies because of their larger inclination corrections. Inclination correction also causes correlated errors because the same inclination is used for both luminosity and velocity width. 

To avoid the problems caused by individual inclination correction, one can model the inclination statistically by treating it as a latent variable with a known probability density function (pdf). It is a prime time to develop such methods, because wide-area \HI\ surveys like ALFALFA \citep{Haynes18} and WALLABY \citep{Westmeier22} continue to produce large samples of \HI-selected galaxies at all inclination angles. In \citet{Fu24}, I introduced an iterative latent-inclination method to restore the TFR from the full ALFALFA sample using measurements {\it not} corrected for inclination. The method is highly efficient and produces a tighter correlation than that from inclination-corrected data. But the method requires binning and the biases in the data are carried directly into the restored TFR. For unbiased inference of the TFR, one could  incorporate the latent variable in the data likelihood function. \citet{Obreschkow13} proposed a latent-inclination MLE method for the inverse TFR, where luminosity is used to predict velocity width. But they missed an important term in their derivation: the intrinsic distribution function of the independent variable. Even though this term drops out in the likelihood function of the inverse TFR (see \S\,\ref{sec:iTFR}), its omission led the authors to incorrectly conclude that for a dual-scatter model, the scatter of the independent variable can be added in quadrature to the scatter of the dependent variable. As a result, their latent-variable MLE model is unidirectional, hence does not account for the general Eddington bias. 

To summarize, precise and unbiased inference of the TFR is limited by the distance-dependent Malmquist bias, the general Eddington bias, and elusive measurement errors in the inclination angle. While the literature addresses each of these problems individually, no solution currently tackles all three simultaneously. For example, the unidirectional MLE method \citep{Willick97} accounts for the distance-dependent Malmquist bias, but it neglects the general Eddington bias from the scatter of the independent variable. The Gaussian-mixture dual-scatter Bayesian method \citep{Kelly07, Bartlett23} can account for the general Eddington bias, but it does not correct for the distance-dependent Malmquist bias because distance is not separated from luminosity in these general regression methods. In addition, neither methods address the inclination-angle problem: they both require the input line widths to be individually corrected by inclination, so low-inclination galaxies are excluded to avoid the most problematic corrections. 
Here I develop latent-variable, dual-scatter, Bayesian methods that incorporate (1) the Sine pdf of inclination angle, (2) the distance-dependent selection function of luminosity, (3) the Gaussian measurement errors and intrinsic scatters in both the dependent and the independent variables, and (4) the distribution function of the independent variable. 
The precision and efficacy of the methods in mitigating the biases will be tested with synthetic data that realistically simulate (1) the luminosity function of galaxies, (2) the distribution in inclination angle and redshift, (3) the data censorship due to detection limit, and (4) the measurement errors and intrinsic scatters in both luminosity and rotation velocity. 

This paper is organized as follows. 
In \S\,\ref{sec:notation}, I describe the observational data used in TFR studies and formulate the TFR problem with shorthand notations.
Next in \S\,\ref{sec:bias}, I start by discussing how biases of the regression coefficient impact measurements of the Hubble constant. Then I describe how observational selection effects and luminosity scatter bias the mean luminosity via the distance-dependent Malmquist bias. Finally, I show how the scatter in the independent variable distorts the distribution of the dependent variable through a formula analogous to the classic Eddington bias of the true mean value. Both biases shift the first moment of the dependent variable away from the true correlation, causing biases in the regression coefficients. 
The next four sections detail the likelihood-based framework that tackles the three primary challenges of Tully-Fisher relation inference: distance-dependent Malmquist bias, general Eddington bias, and inclination correction. First in \S\,\ref{sec:method}, I outline the methodology. Next in \S\,\ref{sec:likefun}, I derive the data likelihood functions for two latent-inclination unidirectional models: the forward model and the inverse model that use rotation velocity and luminosity as the independent variable, respectively. In \S\,\ref{sec:test}, I implement the likelihood functions in a Bayesian framework and employ a Markov-Chain Monte Carlo method to sample the posterior pdf of the inferred parameters. Simulated data of disk galaxies with random sky orientations are then generated to test the numerical methods, to quantify the statistical uncertainties, and to expose the biases of the inferred parameters. As expected, results are biased when the independent variable is measured with error, due to the general Eddington bias described in \S\,\ref{sec:bias}. Two extensions of the unidirectional models are then introduced to mitigate the general Eddington bias. First, in \S\,\ref{sec:iTFR_Edd}, I reverse the expected Eddington bias in the velocity width data and use the moment-shifted data to constrain the inverse model; and in \S\,\ref{sec:uTFR}, I introduce the latent-inclination bidirectional dual-scatter model by incorporating scatter of the independent variable in the likelihood function. Tests on simulated datasets show that both methods yield nearly unbiased estimates of model parameters, confirming the theory in \S\,\ref{sec:bias}. 
Finally, I summarize the work and discuss future extensions and applications in \S\,\ref{sec:summary}. 

To keep the main text focused on key concepts and results, I defer extended equations and important numerical considerations to the Appendices. Appendix\,\ref{sec:step_fun} presents the complete analytical expressions for the conditional pdfs that constitute the likelihood functions for flux-limited samples. Appendix\,\ref{sec:unify_iTFR} gives the alternative derivation of the dual-scatter model, starting from the inverse model, showing that it is bidirectional (symmetric). Appendix\,\ref{sec:accelerate} outlines the key numerical methods that enhanced the evaluation speed of the dual-scatter model's likelihood function by three orders of magnitude, rendering it computationally practical.

In general, Roman letters are used to denote observable quantities and Greek letters are reserved for model parameters. I will use $\tilde{x}$ to indicate measured value (with errors) and $x$ to indicate the true value. Similarly, $\tilde{\theta}$ and $\theta$ represent the estimated value and the true value of parameter $\theta$, respectively. And the bias of the parameter estimation is defined as $B_\theta \equiv \tilde{\theta} - \theta$. 

Python functions and a Jupyter notebook implementing the methods described in this paper are publicly available on GitHub under the MIT license: \url{https://github.com/fuhaiastro/TFR_biases} and on Zenodo under an open-source Creative Commons Attribution license: \dataset[doi:10.5281/zenodo.16378199]{https://doi.org/10.5281/zenodo.16378199}.

\section{Formulation of the Problem} \label{sec:notation}

The TFR has various empirical forms. The luminosity axis could be absolute magnitude, stellar mass, or baryonic mass; and the velocity axis could be line widths from neutral hydrogen (\HI), molecular gas (CO), or ionized gas (\OII). I will formulate the problem using the TFR between baryonic mass and \HI\ line width, because (1) the addition of gas mass to stellar mass tends to reduce the curvature of the correlation \citep[e.g.,][]{McGaugh00, Kourkchi22} and (2) \HI\ is the best tracer of the maximum rotation velocity thanks to its extended spatial distribution. This choice does not restrict the application of the methods to only baryonic TFR, since the derived expressions can be easily adjusted to suit magnitude-based TFRs.

The baryonic TFR is a power-law correlation between the baryonic mass (stellar + gas) and the edge-on rotation velocity. In logarithmic, it is a linear relation with two coefficients ($\beta, \gamma$):
\begin{equation} \label{eq:TFR}
\log M_b = \beta [\log W - \log \sin ({\rm inc}) - 2.5] + \gamma 
\end{equation}
where $\log M_b$ is the logarithmic baryonic mass in solar mass ($M_\odot$) and $\log W - \log \sin ({\rm inc})$ is the observed velocity width ($W$ in \kms) corrected for the inclination angle (${\rm inc}$). The relation is anchored at $\log V_0 = 2.5$\,dex (or $V_0 = 316$\,\kms), so that the intercept $\gamma$ is the baryonic mass of spirals with edge-on velocity width of 316\,\kms. 

The baryonic mass is calculated as the {\it apparent} baryonic mass multiplied by the square of the luminosity distance, which in logarithmic is: 
\begin{equation}
\log M_b = \log m_b + 2 \log D_L
\end{equation}
The apparent mass ($\log m_b$) includes the $4\pi$ term and is calculated using the \HI\ 21\,cm flux ($S_{21}$ in Jy\,km\,s$^{-1}$) and apparent magnitude in an optical/near-IR filter ($m_\lambda$) using the following relation: 
\begin{equation} \label{eq:app_mass}
m_b = 2.356\times10^5 K_g S_{21} + 10^{-0.4(m_\lambda-M_{\lambda,\odot}-25)} (M/L)_\lambda
\end{equation} 
where $K_g$ is the \HI-to-gas conversion factor (typically assumed to be 1.33 to account for Helium), $M_{\lambda,\odot}$ is the absolute magnitude of the Sun \citep{Willmer18}, and $(M/L)_\lambda$ is the color-estimated mass-to-light ratio in solar units ($M_\odot/L_\odot$). The latter two must be in the same filter as the galaxy magnitude, as indicated by the subscript $\lambda$. 

For most galaxies, distance is not a direct observable, but rather an inferred quantity from the observed redshift and a cosmological model. At low redshift ($z < 0.1$), the redshift-derived luminosity distance can be accurately calculated using the Taylor expansion form of \citet{Caldwell04}:
\begin{equation} \label{eq:DL}
D_L \approx \frac{cz}{H_0} \left[1+\frac{1}{2}(1-q_0)z - \frac{1}{6} (1-q_0-3q_0^2+j_0) z^2 \right]
\end{equation}
where $cz$ is the cosmological redshift in \kms, $H_0$ = 70\,km\,s$^{-1}$\,Mpc$^{-1}$ is the assumed Hubble constant, $q_0 = -\frac{\ddot{a}a}{\dot{a}^2} = \Omega_{m,0}/2-\Omega_{\Lambda,0}$ is the deceleration parameter, and $j_0 = -\frac{\dddot{a}a^2}{\dot{a}^3} = \Omega_{m,0}+\Omega_{\Lambda,0}$ is the cosmic jerk. When adopting $\Omega_{m,0} = 0.315, \Omega_{\Lambda,0} = 0.685$ from \citet{Planck-Collaboration20VI}, $q_0 = -0.53$ and $j_0 = 1$. 

In the following, I will use shorthand symbols for the logarithmic of the apparent mass, the distance squared, the velocity width ratio, and the sine of the inclination: 
\begin{align} \label{eq:shorthands}
m &\equiv \log m_b \nonumber \\
d &\equiv 2 \log D_L \nonumber \\
w &\equiv \log W-2.5 \nonumber \\
i &\equiv \log \sin ({\rm inc})
\end{align}
With these shorthands, the TFR in Eq.\,\ref{eq:TFR} can be rewritten as:
\begin{equation} \label{eq:TFRshort}
m+d = \beta(w-i) + \gamma 
\end{equation}

The data available to the astronomer are $\{\tilde{m}_i,\tilde{w}_i,\tilde{d}_i\}_{k=1}^N$ for a sample of $N$ galaxies that survived the observational selection function $S(\tilde{m},\tilde{w})$. The goal of parameter inference is to optimally utilize these data to constrain the TF parameters $(\beta, \gamma)$, ideally free of statistical and selection biases. 

\section{Selection Bias and Statistical Bias} \label{sec:bias}

It is important to mitigate biases in the inferred regression coefficients of the TFR because a biased $\gamma$ directly leads to a biased Hubble constant ($H_0$). Two separate samples are needed to estimate $H_0$: a larger and more distant redshift sample and a smaller and closer zero-point sample. The distances of the former are calculated using an assumed value of $H_0$, and the distances of the latter are from standard candles on the lower rung of the distance ladder (e.g., Cepheids). The two separate samples are needed for $H_0$ measurement because (1) the redshifts of the zero-point sample are strongly influenced by peculiar velocities, and (2) the distances of the redshift sample are inferred from an assumed $H_0$. The TF coefficients ($\beta,\gamma$) are estimated using the redshift sample, while the zero-point sample determines the intercept ($\gamma_{ZP}$) with Cepheid distances assuming the same slope $\beta$. Since it is fair to assume that both samples follow the same TFR, any difference between the intercepts from the two samples ($\gamma$ and $\gamma_{ZP}$) implies a deviation between the actual value of $H_0$ and the assumed Hubble constant (70\,\hunit):
\begin{equation}
\log(H_0/70) = 0.5 (\gamma - \gamma_{ZP})
\end{equation} 
As a result, the bias in $\gamma$ will directly transfer to the inferred Hubble constant: 
\begin{equation} \label{eq:H0_bias}
B_{H_0} \approx 35 \ln 10 B_\gamma = 4.84 (B_\gamma/0.06)\,{\rm km}\,{\rm s}^{-1}\,{\rm Mpc}^{-1}
\end{equation}
A bias of 0.06\,dex (or 0.15\,mag) in the intercept would lead to a bias of $\sim$5\,\hunit\ in the Hubble constant, comparable to the current ``Hubble tension'' \citep[e.g.][]{Riess24}.

Note that $B_\gamma$ is only one of the two biases involved in the Hubble constant; the other is the bias of $\gamma_{ZP}$ from the zero-point sample. It is thus critical to use the same statistical model for the redshift and the zero-point samples, in addition to consistent measurements of rotation velocities and masses (or luminosities) across the two samples \citep{Bradford16}. If the intercept biases from the two samples are equal, they will cancel out, resulting in an unbiased $H_0$ estimate. However, this is unlikely, as the biases depend on sample characteristics that typically differ between the two samples. Therefore, to obtain an unbiased Hubble constant determination, both samples must be analyzed using statistical models that account for their specific characteristics to correct for their distinct biases. In the following, I describe how random Gaussian scatter in both the dependent and independent variables biases regression coefficients through mechanisms akin to the classic Malmquist and Eddington biases.

The classic Malmquist bias \citep{Eddington14, Malmquist22} is caused by data censorship and the Gaussian luminosity scatter of a standard candle. The classic Eddington bias of the mean true value \citep{Dyson26, Eddington40} is due to measurement error and gradient in the distribution function of the measurements. The two biases are orthogonal to each other: Malmquist takes the true mean luminosity as known and compare it to the mean of the measured luminosities that survived the selection ($BM \equiv \langle \tilde{x} \rangle - x$), while Eddington takes the measurement as known and compare it to the mean of the unknown true values ($BE \equiv \tilde{x} - \langle x \rangle$). However, the two biases are often conflated because, in the simplest cases, their mathematical outcomes appear similar. The classic Malmquist bias of absolute magnitude for uniformly distributed standard candles with a Gaussian luminosity function of a mean of $M$ and a standard deviation of $\sigma_M$ is $\langle \tilde{M} \rangle - M = -1.382\sigma_M^2$, and the Eddington bias of apparent magnitude for uniformly distributed sources with constant luminosity and a measurement error of $\sigma_m$ is $\tilde{m} - \langle m \rangle = -1.382\sigma_m^2$. In both cases, the constant 1.382 is $0.6 \ln 10$. 

To formulate the problem, consider a family of standardizable candles on an idealized linear correlation: $y(x) = \beta x + \gamma$. For galaxies on the TFR, $y$ would be the logarithmic luminosity and $x$ the inclination-corrected logarithmic velocity width. And the inverse function of $y(x)$ is $x(y) = (y-\gamma)/\beta$.

First, consider random Gaussian scatter only in the luminosity axis, $\tilde{y} = y(x) + \epsilon_y$, where the random variable $\epsilon_y$ is drawn from a normal distribution with zero mean and a standard deviation of $\sigma_y$. So the conditional pdf of $\tilde{y}$ at a given $x$ is a Gaussian:
\begin{equation}
p(\tilde{y}|x) = \frac{1}{\sqrt{2\pi}\sigma_y} \exp \left(-\frac{[\tilde{y}-y(x)]^2}{2\sigma_y^2}\right)
\end{equation}
When the sample is selected above a logarithmic luminosity limit of $y_l$ that depends on distance ($D_L$; $y_l = 2 \log D_L + {\rm const.}$ for a flux-limited sample), the selection function and the luminosity dispersion $\sigma_y$ shifts the mean luminosity of the observed sample relative to the correlation-predicted mean luminosity according to:
\begin{align} \label{eq:Malm_dist}
y(x) - \langle \tilde{y} \rangle_x 
&= \frac{\int_{y_l}^{\infty} [y(x)-\tilde{y}] p(\tilde{y}|x) d\tilde{y}}{\int_{y_l}^{\infty} p(\tilde{y}|x) d\tilde{y}} \nonumber \\
&= -\sigma_y \sqrt{\frac{2}{\pi}} \frac{\exp [-(y_l - y(x))^2/(2\sigma_y^2)]}{{\rm erfc} [(y_l - y(x))/(\sqrt{2}\sigma_y)]} 
\end{align}
For the TFR, $y(x)$ is the correlation-predicted luminosity for a given rotation velocity and $\langle \tilde{y} \rangle_x$ is the mean of  measured luminosities at the same rotation velocity. Because the difference between the two depends on distance, this is termed the ``distance-dependent Malmquist bias'' \citep[e.g.,][]{Willick94, Butkevich05}.

Next, consider the scatter is instead only in the $x$-axis, $\tilde{x} = x(y) + \epsilon_x$, where the random variable $\epsilon_x$ is drawn from a normal distribution with zero mean and a standard deviation of $\sigma_x$. In this case, there is no selection function, but the mean $y$ will still be shifted from the correlation-predicted value, because the distribution of $y$ at a given $\tilde{x}$ is altered from a Gaussian according to Bayes' rule:
\begin{equation}
p(y|\tilde{x}) = \frac{p(\tilde{x}|y) p(y)}{p(\tilde{x})}
\end{equation}
where the conditional pdf of $\tilde{x}$ at a given $y$ is a Gaussian:
\begin{align}
p(\tilde{x}|y) &= \frac{1}{\sqrt{2\pi}\sigma_x}  \exp \left(-\frac{[\tilde{x}-x(y)]^2}{2\sigma_x^2}\right) \nonumber \\
	&= \frac{1}{\sqrt{2\pi}\sigma_x} \exp \left(-\frac{[y(\tilde{x})-y]^2}{2(\beta\sigma_x)^2} \right)
\end{align}

For the TFR, the difference between the correlation-predicted luminosity at a measured rotation velocity [$y(\tilde{x})$] and the mean luminosity at the same rotation velocity ($\langle y \rangle_{\tilde{x}}$) can be calculated as:
\begin{align} 
y(\tilde{x}) &- \langle y \rangle_{\tilde{x}} = \int [y(\tilde{x})-y] \frac{p(\tilde{x}|y) p(y)}{p(\tilde{x})} dy \nonumber \\
&= \int [y(\tilde{x})-y] \frac{1}{\sqrt{2\pi}\sigma_x} \exp \left(-\frac{[y(\tilde{x})-y]^2}{2(\beta\sigma_x)^2} \right) \frac{p(y)}{p(\tilde{x})} dy  \nonumber \\
&= \frac{\beta^2\sigma_x}{\sqrt{2\pi} p(\tilde{x})} \int \frac{\partial}{\partial y} \left[ \exp \left(-\frac{[y(\tilde{x})-y]^2}{2(\beta\sigma_x)^2} \right) \right] p(y) dy 
\end{align}
The symmetry between $y$ and $y(\tilde{x})$ in the exponential function allows the chain rule: 
\begin{align} 
&y(\tilde{x}) - \langle y \rangle_{\tilde{x}} \nonumber \\
&= -\frac{\beta^2\sigma_x}{\sqrt{2\pi} p(\tilde{x})} \int \frac{\partial}{\partial y(\tilde{x})} \left[ \exp \left(-\frac{[y(\tilde{x})-y]^2}{2(\beta\sigma_x)^2} \right) \right] p(y) dy  \nonumber \\
&= -\frac{\beta\sigma_x^2}{p(\tilde{x})} \int \frac{\partial}{\partial\tilde{x}} \left[p(\tilde{x}|y) p(y)\right] dy
\end{align}
The final result is obtained by taking the derivative out of the integral (Leibniz's integral rule):
\begin{align} \label{eq:Edd_gen_y}
y(\tilde{x}) - \langle y \rangle_{\tilde{x}} &= -\frac{\beta\sigma_x^2}{p(\tilde{x})} \frac{d}{d\tilde{x}} \left( \int p(\tilde{x}|y) p(y) dy \right) \nonumber \\
&= -\beta\sigma_x^2 \frac{d \ln p(\tilde{x})}{d\tilde{x}} 
\end{align}

The result is analogous to the classic Eddington bias of the mean true value \citep{Dyson26, Eddington40}:
\begin{equation} \label{eq:Edd}
\tilde{x} - \langle x \rangle_{\tilde{x}} = -\sigma^2 \frac{d \ln p(\tilde{x})}{d \tilde{x}}
\end{equation}
where $\tilde{x}$ is the measured value, $\langle x \rangle_{\tilde{x}} = \int x p(x|\tilde{x}) dx$ the mean of the true values that could have produced the measured value, $\sigma$ the measurement error, and $p(\tilde{x})$ the distribution function of the measured values. In fact, the classic Eddington bias is a special case of Eq.\,\ref{eq:Edd_gen_y} for $\beta = 1$ and $\gamma = 0$. Thus, this bias is termed the ``general Eddington bias''.

Both biases affect the estimates of regression coefficients by distorting the underlying data distribution. It is easy to imagine how the general Eddington bias leads to biased regression coefficients. Suppose $p(\tilde{x})$ is a simple exponential distribution function, $p(\tilde{x}) \propto \exp(\tilde{x}/x_0)$, the Eddington bias is a constant, implying a vertical shift of the best-fit relation from the truth, causing the intercept to change while keeping the slope intact. For other distributions (e.g., power-law or Schechter function), the Eddington bias varies with $\tilde{x}$, leading both the slope and the intercept to drift away from the truth. The distance-dependent Malmquist bias affects the regression coefficients in a more complicated way because the bias depends not only on luminosity scatter and flux limit, but also on redshift and the unknown true correlation.

The biases operate on the data simultaneously but are driven by different sample characteristics: Malmquist by luminosity dispersion and luminosity limit [$\sigma_y, y_l(cz)$], while Eddington by dispersion of the independent variable and its distribution function [$\sigma_x, p(\tilde{x})$]. From a modeling perspective, correcting the biases in parameter estimation thus requires statistical models that build in all of those sample characteristics [$\sigma_y, y_l(cz), \sigma_x, p(\tilde{x})$].

\section{Overview of the Methodology} \label{sec:method}
 
Although the TFR is often described as a simple linear correlation in logarithmic scales, the inference of its slope and intercept is not a simple linear regression problem, because of several important characteristics of the TFR problem:
\begin{itemize}
\item The dependent variable (e.g., baryonic mass) is a combination of two observables: the apparent mass and the luminosity distance from redshift. For flux-limited samples, the selection function of the dependent variable depends on distance (not mass).
\item The independent variable (maximum rotation velocity) requires de-projecting the apparent \HI\ line width to the edge-on perspective. So it requires estimating the inclination angle of the disk relative to the line-of-sight. 
\item There are measurement errors in all four observables, and there are intrinsic dispersions in both axes\footnote{The dual intrinsic dispersion model posits that galaxies with identical luminosity may exhibit a range of maximum rotation velocities, and similarly, galaxies with identical rotation velocities may show a range of luminosities. The extent of these dispersions, both absolute and relative, should be derived from the data rather than arbitrarily set by the observer. Like measurement errors, the intrinsic Gaussian dispersions along the two axes cannot be merged into a single Gaussian when distribution functions are non-uniform.}.
\end{itemize}

Because of these characteristics, precise and unbiased inference of the TFR is limited by three major factors:
\begin{itemize}
\item The uncertain estimate of the inclination angle. 
\item The distance-dependent Malmquist bias (Eq.\,\ref{eq:Malm_dist}) due to (1) the separation of flux limit and luminosity limit in a sample covering a range of distances, and (2) the measurement error and the intrinsic dispersion in mass.  
\item The general Eddington bias (Eq.\,\ref{eq:Edd_gen_y}) due to (1) non-uniform distributions of galaxies in rotation velocity, and (2) the measurement error and the intrinsic scatter in rotation velocity.  
\end{itemize}

In this work, I will tackle all three problems simultaneously in a self-consistent Bayesian framework:
\begin{itemize}
\item To avoid errors associated with inclination measurements, galaxy inclination is treated as a latent variable with a known pdf, which is then marginalized when computing the data likelihood function.
\item To mitigate the Malmquist bias, apparent masses and redshift-inferred distances are kept separate in all expressions, and the conditional probability of the apparent mass is normalized to account for the distance-dependent sample truncation.
\item To mitigate the Eddington bias, two approaches are introduced: (1) the analytical formula of the bias from \S\,\ref{sec:bias} is utilized to correct for the moment of the data in an iterative fashion (\S\,\ref{sec:iTFR_Edd}), and (2) both the Schechter distribution function and the scatter of the independent variable are included in the data likelihood function of a ``bidirectional dual-scatter model'' (\S\,\ref{sec:uTFR}). The model is bidirectional (symmetric), producing equivalent likelihood functions regardless of whether rotation velocity or baryonic mass is selected as the independent variable.
\end{itemize}

\section{Derivation of the Likelihood Functions} \label{sec:likefun}

To begin the Bayesian inference process, I derive the data likelihood functions for the two unidirectional models: the forward model with rotation velocity as the independent variable and the inverse model with mass as the independent variable. These models are useful because they isolate the Eddington bias (\S\,\ref{sec:test_result}) and lay the foundation for the more complicated bidirectional dual-scatter model (\S\,\ref{sec:uTFR}).

\subsection{The Forward Model} \label{sec:fTFR}

The forward model uses velocity width $(w-i)$ to predict the mass $(m+d)$. It also assumes Gaussian dispersion along the mass axis. By using $w$ as an independent variable, its measurement error is inherently ignored in the forward model. To make this explicit in the notation and to distinguish the forward model with the dual-scatter model in \S\,\ref{sec:uTFR}, I use $w$ in place of $\tilde{w}$ in this subsection. In addition, the residual redshift noise propagated into the distance parameter $\tilde{d}$ is ignored and $d$ is used in place of $\tilde{d}$ to make this assumption explicit. But I will comment on how to properly include redshift noise in the likelihood function.

Given the TFR in Eq.\,\ref{eq:TFRshort}, the forward model can be expressed as: 
\begin{equation} \label{eq:model_fTFR}
\tilde{m} = \beta (w - i) - d + \gamma + \epsilon_{im} + \epsilon_{em}
\end{equation}
where the random variables represent the intrinsic dispersion ($\epsilon_{im}$) and the measurement error ($\epsilon_{em}$), and they are drawn from Gaussian distributions with zero means and standard deviations of $\sigma_{im}$ and $\sigma_{em}$, respectively.

In addition, the forward model also requires parameterizing the distribution function of edge-on rotation velocity ($w-i$) of the galaxy population (hereafter ``velocity function''). Given the TFR that links velocity with mass, the velocity function is closely related to the mass function, which is parameterized as a Schechter function:
\begin{align} \label{eq:MF}
\frac{dn}{d(m+d)} = \phi_\star 10^{(\alpha+1)(m+d-M_\star)} \exp (-10^{m+d-M_\star})
\end{align}
where $\alpha$ is the faint-end slope, $M_\star$ the characteristic mass (above which the exponential drop-off commences). Typically, the normalization factor $\phi_\star$ is in units of Mpc$^{-3}$\,dex$^{-1}$; for the purpose here, I redefine $\phi_\star$ as the factor that normalizes the distribution function to unity so that it becomes a properly normalized pdf $p(m+d)$. 

Substituting $m+d$ with $\beta(w-i)+\gamma$, one obtains the corresponding velocity function: 
\begin{align} \label{eq:VF}
\frac{dn}{d(w-i)} = \beta \phi_\star 10^{(\alpha+1)[\beta (w-i)-v_\star]} \exp [-10^{\beta(w-i)-v_\star}]
\end{align}
where $v_\star = M_\star - \gamma$, which equals the characteristic logarithmic velocity multiplied by $\beta$. Notice the extra $\beta$ in front of $\phi_\star$ because $d(m+d) = \beta d(w-i)$. 

In the following, I will start by expressing the joint probability of $(\tilde{m},d,w,i)$, then marginalize it over $i$ to obtain the joint probability of the observables $(\tilde{m},d,w)$, and finally using the joint probability and the selection function to calculate the conditional probability of $\tilde{m}$, which makes up the likelihood function.

The joint probability that a galaxy has the three observables $(\tilde{m}, \tilde{d}, w)$ and an inclination $i$ can be factorized using the multiplication rule:
\begin{equation} \label{eq:joint_fTFR}
p(\tilde{m},d,w,i) = p(\tilde{m}|w,i,d) p(w|i) p(i) p(d)
\end{equation}
When choosing these factors, I have considered relations between the variables and have made the following assumptions:
\begin{itemize}
\item the velocity function parameters does not evolve over the redshift range covered by the sample, so that $p(w|i)$ does not depend on distance.
\item inclination and distance are independent variables, so that $p(i,d) = p(i) p(d)$.
\end{itemize}

The particular set of conditional pdfs are straightforward to express. The first term is obtained by marginalizing the join pdf $p(\tilde{m},m|w,i,d)$ over $m$, which is a convolution of two Gaussians:
\begin{align} \label{eq:gaus_m}
p(\tilde{m}|w,i,d) &= \int_{-\infty}^{\infty} p(\tilde{m}|m) p(m|w,i,d) dm \nonumber \\
& = \frac{1}{{\sigma_m \sqrt {2\pi } }} \exp \left[ -\frac{[\tilde{m}+d-\beta(w-i)-\gamma]^2 }{2\sigma_m^2 } \right]
\end{align}
Per the convolution theorem of Gaussians, the variance of the convolved Gaussian, $\sigma_m^2$, is the sum of the variances from intrinsic dispersion and measurement error:
\begin{equation} \label{eq:sigm}
\sigma_m^2 = \sigma_{im}^2 + \sigma_{em}^2
\end{equation}
Note that the likelihood-based method naturally handles heteroscedastic measurement errors, because a different $\sigma_{em}$ can be supplied for each source while $\sigma_{im}$ is kept as a constant for the whole sample. 

The second term is the probability of projected velocity given the inclination and it can be expressed by the normalized Schechter velocity function in Eq.\,\ref{eq:VF}: 
\begin{equation} \label{eq:velfun}
p(w|i) = \beta \phi_\star 10^{(\alpha+1) [\beta (w-i)-v_\star]} \exp [-10^{\beta(w-i)-v_\star}]
\end{equation}
The velocity function is a critical component of the forward model because the term depends on $i$ so it does not drop out in the conditional pdf in Eq.\,\ref{eq:cpdf_fTFR} below. 

The third term is the prior pdf of the inclination parameter. Assuming isotropic random orientation on the sky, the pdf of inclination angle ${\rm inc}$ is a sine function: $p({\rm inc}) = \sin ({\rm inc})$ for $0 < {\rm inc} < \pi/2$. The pdf of $i \equiv \log \sin ({\rm inc})$ can then be derived using the pdf identity $p(i) di = p({\rm inc}) d{\rm inc}$, and the result is:
\begin{equation} \label{eq:pi}
p(i) = \ln 10 \frac{10^{2i}}{\sqrt{1-10^{2i}}}
\end{equation}
Note that $i \leq 0$ given its definition. 

The last term is the probability of observing a galaxy at a given distance. It is proportional to the integrated galaxy volume density at the distance $n(d)$ (i.e., the redshift distribution) multiplied by the survey volume $\Omega D_L^2 d D_L$. Expressed in distance parameter $d$, the probability is:
\begin{equation} \label{eq:pd}
p(d) \propto 10^{3d/2} n(d)
\end{equation}
When one ignores the difference between the redshift-inferred distance $\tilde{d}$ and the true distance $d$ (as assumed here), this term drops out in the conditional pdf in Eq.\,\ref{eq:cpdf_fTFR} below. But if one opts to account for this difference \citep[e.g.,][]{Willick97}, the following term needs to be multiplied to the joint pdf in Eq.\,\ref{eq:joint_fTFR2} below to obtain $p(\tilde{m},w,\tilde{d},d,i)$:
\begin{equation}
p(\tilde{d} | d) = \frac{1}{{\sigma_d \sqrt {2\pi } }} \exp \left[ -\frac{(\tilde{d}-d)^2}{2\sigma_d^2 } \right]
\end{equation}
where the standard deviation is proportional to the ratio between the velocity noise ($\sigma_{cz}$) and the redshift:
\begin{equation}
\sigma_d \approx \frac{2}{\ln 10} \frac{\sigma_{cz}}{cz}
\end{equation}
and an additional marginalization over $d$ is required when calculating the joint pdf of the three observables in Eq.\,\ref{eq:marg_fTFR}. Note that because of the additional $d$-dependent terms in $p(d)$ (Eq.\,\ref{eq:pd}), the marginalization over $d$ is not as simple as the marginalization over $m$ in Eq.\,\ref{eq:gaus_m}. For this reason, it is incorrect to absorb this integral over $d$ by simply adding $\sigma_d^2$ to $\sigma_m^2$ in Eq.\,\ref{eq:sigm}.

Combining the terms defined above, the joint probability can be written is:
\begin{align} \label{eq:joint_fTFR2}
p(\tilde{m},w,d,i) &= p(d) \times \frac{\ln 10 \cdot 10^{2 i }}{\sqrt{1-10^{2 i}}} \nonumber \\
&\times \frac{1}{{\sigma_m \sqrt {2\pi } }} \exp \left[ -\frac{[\tilde{m}+d-\beta(w-i)-\gamma]^2 }{2\sigma_m^2 } \right] \nonumber \\
&\times \beta \phi_\star 10^{(\alpha+1) [\beta (w-i)-v_\star]} \exp [-10^{\beta(w-i)-v_\star}]
\end{align}
which can then be marginalized over $i$ to obtain the joint probability of the three observables:
\begin{equation} \label{eq:marg_fTFR}
p(\tilde{m},w,d) 
= \int_{-\infty}^0 p(\tilde{m},w,d,i) \, di
\end{equation}

The observational selection function, $S(\tilde{m},w)$, truncates the joint pdf, $p(\tilde{m},w,d)$, so that the resulting pdf is no longer normalized. To account for $S(\tilde{m},w)$ and the associated Malmquist bias, one shall compute the conditional pdf of $\tilde{m}$ by normalizing the truncated joint pdf: 
\begin{align} \label{eq:cpdf_fTFR}
p(\tilde{m} &| w, d) = 
\frac{S(\tilde{m},w) p(\tilde{m}, w, d)}{\int_{-\infty}^{\infty} S(\tilde{m},w) p(\tilde{m}, w, d) \, d\tilde{m}} \nonumber \\
&= \frac{S(\tilde{m},w) \int_{-\infty}^0 p(\tilde{m}|w,i,d) p(w|i) p(i) \, di}{\int_{-\infty}^{\infty} S(\tilde{m},w) \int_{-\infty}^0 p(\tilde{m}|w,i,d) p(w|i) p(i) \, di \, d\tilde{m}} 
\end{align}
Notice that the distance distribution function, $p(d)$, has cancelled out in the division. When the sample is simply flux-limited by a step function, the integral over $\tilde{m}$ on the denominator becomes an error function, making it much faster to evaluate. I provide the full expression of the conditional pdf for flux-limited samples in Eq.\,\ref{eq:cpdf_fTFR_step} in Appendix\,\ref{sec:step_fun}. 

The data likelihood function ($\mathcal{L}$) is defined as the probability of the data given the model. For the forward model, the data are $\{\tilde{m}_k\}_{k=1}^N$ and the model is the combination of the model parameters $\boldsymbol\theta \equiv (\beta,\gamma,\sigma_m,v_\star,\alpha)$ and the independent variables $\{w_k,d_k\}_{k=1}^N$. It is thus the product of the conditional probabilities in Eq.\,\ref{eq:cpdf_fTFR} of all valid data points. In practice, its logarithmic is preferred:
\begin{equation} \label{eq:lnL_fTFR}
\ln {\mathcal L} = \sum_{k=1}^N \ln p(\tilde{m}_k | w_k, d_k)
\end{equation} 

\subsection{The Inverse Model} \label{sec:iTFR}

The inverse model uses mass ($m+d$) to predict velocity width ($w-i$) and assumes Gaussian dispersion along the velocity axis: 
\begin{equation}
\tilde{w}-i = (m + d - \gamma)/\beta + \epsilon_{iw} + \epsilon_{ew}
\end{equation}
where the random variables represent the intrinsic dispersion in velocity ($\epsilon_{iw}$) and measurement error ($\epsilon_{ew}$). By using $m+d$ as the independent variable, its measurement error is inherently ignored in the inverse model, so I use $m$ and $d$ in place of $\tilde{m}$ and $\tilde{d}$ in this subsection.

Starting by factorizing the joint pdf:
\begin{equation} \label{eq:joint_iTFR}
p(\tilde{w},m,d,i) = p(\tilde{w}|m,d,i) p(m|d) p(d) p(i)
\end{equation} 
Similar to the forward model, the first term is obtained by marginalizing the joint pdf $p(\tilde{w},w|m,d,i)$ over $w$:
\begin{align} \label{eq:gaus_w}
p(\tilde{w}|m,d,i) &= \int_{-\infty}^{\infty} p(\tilde{w}|w) p(w|m,d,i) dw \nonumber \\
&= \frac{1}{{\sigma_w\sqrt{2\pi}}} \exp \left[ -\frac{[\tilde{w}-i-(m+d-\gamma)/\beta]^2}{2\sigma_w^2} \right]
\end{align}
where the variance $\sigma_w^2$ is the sum of the variances from intrinsic dispersion and measurement error: 
\begin{equation} 
\sigma_w^2 = \sigma_{iw}^2 + \sigma_{ew}^2
\end{equation}  
The second term is the normalized mass function in Eq\,\ref{eq:MF}:
\begin{equation} \label{eq:pmd_iTFR}
p(m|d) = \phi_\star 10^{(\alpha+1)(m+d-M_\star)} \exp (-10^{m+d-M_\star})
\end{equation} 
For simplicity, I assume that the baryonic mass function is independent of inclination angle, implying that intrinsic dust extinction has been corrected for when calculating the stellar-mass-to-light ratio using color. The last two terms of the joint pdf are the same as in Eqs.\,\ref{eq:pi} and \ref{eq:pd}. Therefore, the joint pdf can be written as:
\begin{align} \label{eq:joint_iTFR2}
p(\tilde{w},m,d,i) &= p(d) \times \frac{\ln 10 \cdot 10^{2 i }}{\sqrt{1-10^{2 i}}} \nonumber \\
&\times \frac{1}{{\sigma_w \sqrt {2\pi } }} \exp \left[ -\frac{[\tilde{w}-i-(m+d-\gamma)/\beta]^2 }{2\sigma_w^2 } \right] \nonumber \\
&\times \phi_\star 10^{(\alpha+1) (m+d-M_\star)} \exp (-10^{m+d-M_\star})
\end{align}
which can then be marginalized over $i$ to obtain the joint probability of the three observables:
\begin{equation}
p(\tilde{w},m,d) = \int_{-\infty}^0 p(\tilde{w},m,d,i) di
\end{equation}

The conditional probability of $\tilde{w}$ is then obtained by normalization:
\begin{align} \label{eq:cpdf_iTFR}
p(\tilde{w}|m,d) 
&= \frac{S(m,\tilde{w}) p(\tilde{w}, m, d)}{\int_{-\infty}^{\infty} S(m,\tilde{w}) p(\tilde{w}, m, d) \, d\tilde{w}} \nonumber \\  
&= \frac{S(m,\tilde{w}) \int_{-\infty}^0 p(\tilde{w}|m,i,d) p(i) \, di}{\int_{-\infty}^{\infty} S(m,\tilde{w}) \int_{-\infty}^0 p(\tilde{w}|m,i,d) p(i) \, di \, d\tilde{w}}
\end{align}
where both the mass distribution function, $p(m|d)$, and the distance distribution, $p(d)$, have dropped out. Hence, only the three TFR model parameters are constrained by the inverse model: $\boldsymbol\theta \equiv (\beta, \gamma, \sigma_w)$. When the selection function depends only on $m$ [i.e., $S(m,\tilde{w}) = S(m)$], it also drops out from the expression, further simplifying the result:
\begin{align} \label{eq:cpdf_iTFR_Sm}
&p(\tilde{w}|m,d) = \int_{-\infty}^0 p(\tilde{w}|m,d,i) p(i) \, di \nonumber \\
&= \frac{\ln 10}{{\sigma_w\sqrt{2\pi}}} \int_{-\infty}^0 \frac{10^{2 i }}{\sqrt{1-10^{2 i}}}
\exp \left[ -\frac{[\tilde{w}-i-(m+d-\gamma)/\beta]^2}{2\sigma_w^2} \right] \, di
\end{align}  

The data likelihood function is obtained from the conditional pdf in the same manner as in the forward model:
\begin{equation} \label{eq:lnL_iTFR}
\ln {\mathcal L} = \sum_{k=1}^N \ln p(\tilde{w}_k | m_k, d_k)
\end{equation} 

\section{Implementation and Testing} \label{sec:test}

This section describes how I implement the latent-inclination unidirectional models in a Bayesian framework (\S\,\ref{sec:mcmc}),  simulate TF galaxy samples with random sky orientations (\S\,\ref{sec:simulate}), and test the models with simulated data sets (\S\,\ref{sec:test_result}). The testing reveals the biases of inferred parameters when the independent variable randomly scatters, as expected from the general Eddington bias described in \S\,\ref{sec:bias}. For measurement errors and sample sizes typical for the ALFALFA sample, the biases are much greater than the inferred statistical uncertainties and they scale with the scatter of the independent variable.

\subsection{Parameter Inference} \label{sec:mcmc}

The Bayesian theorem uses the data likelihood function $\ln \mathcal{L}$ to update the prior knowledge of the model to provide the posterior distribution of the model parameters given the data ($D$):
\begin{equation} \label{eq:post_fTFR}
\ln p(\boldsymbol\theta | D) = \ln \mathcal{L} + \ln p(\boldsymbol\theta) - \ln p(D)
\end{equation}
where $p(\boldsymbol\theta)$ is the prior pdf of the model parameters and $p(D)$ is the Bayesian evidence of the model as a whole. 

When prior knowledge is insufficient or ignored, the priors are assumed to be flat within specified bounds:
\begin{equation}
p(\boldsymbol\theta) = \prod_j \frac{1}{\Delta \theta_j}
\end{equation} 
When such bounded flat priors are assumed, the posterior peaks at the parameters that maximizing the data likelihood, making the Bayesian inference essentially the same as MLE.

The evidence is defined as the marginalized probability of the data ($D$) over all model parameters ($\boldsymbol \theta$) of a given model:
\begin{equation}
p(D) = \int p(D|\boldsymbol\theta) p(\boldsymbol\theta) d\boldsymbol\theta
\end{equation}

It is often challenging to evaluate the evidence accurately because it is a multi-dimensional integral that requires calculations of the data likelihood function over the entire parameter space. But it can be safely ignored for the purpose of parameter inference, because it is a constant for a given model and a given data set. In addition, Markov-Chain Monte Carlo (MCMC) algorithms do not require {\it normalized} pdf to sample the parameter space at frequencies proportional to the pdf, making them the favored method for Bayesian inference. 

I employ the affine-invariant MCMC ensemble sampler implemented in the Python code \texttt{emcee}\footnote{\url{https://emcee.readthedocs.io}} \citep{Foreman-Mackey13} to sample the posterior pdfs. The ensemble approach is naturally parallelizable, inherently handles correlated parameters, and minimizes the need for manual tuning of step sizes for poorly scaled parameters. Standard procedure to set up the sampler is followed. The number of walkers is set to an integer number of CPU cores and is set between two and three times the number of free parameters. Starting from random initial positions within the bounds, the walkers proceed until the length of the chains exceeds 50 times the estimated autocorrelation length ($l$; typically between 40 and 200 steps). After discarding the initial $2l$ steps (``burn-in'') and thinning the chains by keeping one step for every $l/2$ steps, the chains from all walkers are combined to form the final parameter array, whose distribution functions should trace the posterior pdfs of the parameters given the data. For each model parameter, I quote the best-fit value and the $1\sigma$ (68\%) credible interval given by the 50th, 16th, and 84th percentiles of the marginalized cumulative distribution function. 

\subsection{Simulated Data for Testing} \label{sec:simulate} 

\begin{figure*}[!tb]
\epsscale{1.0}
\plotone{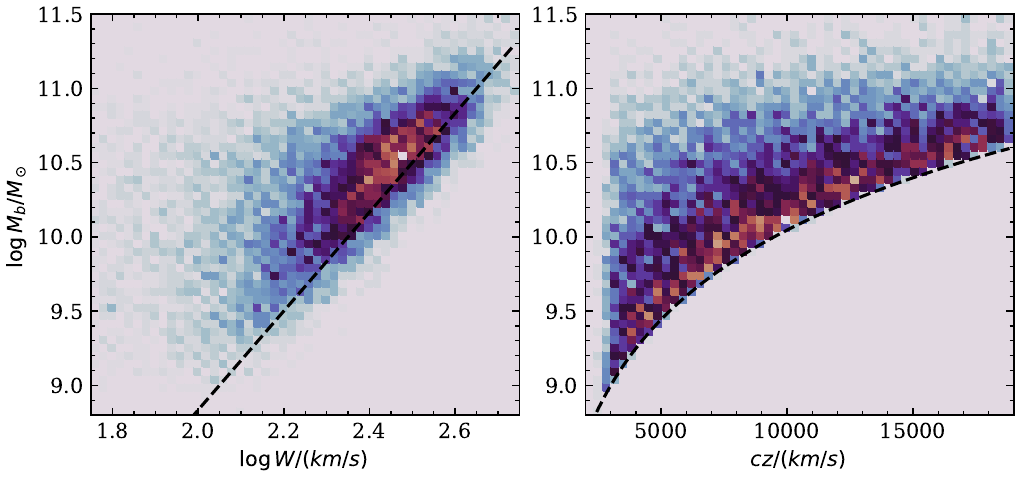}
\caption{Distributions of simulated sample C in baryonic mass vs. projected line width (left) and baryonic mass vs. redshift (right). The dashed line in the left panel shows the prescribed TFR and the dashed curve in the right panel shows the mass limit as a function of redshift.  
\label{fig:data_dist}} 
\epsscale{1.0}
\end{figure*}

\begin{table}
\begin{center}
\caption{Input Parameters vs. Inferred Parameters}
\label{tab:pars}
\begin{tabular}{lrrr}
\hline
\hline
 & \multicolumn{3}{c}{Simulated Data Set} \\
Parameter & A & B & C  \\
\hline
\multicolumn{4}{l}{Input Parameters (Truth)} \\
\hline
$\sigma_m$ & 0.15 & 0.0 & 0.15 \\
$\sigma_w$ & 0.0 & 0.045 & 0.045 \\
$\gamma$ & 10.50 & 10.50 & 10.50 \\
$\beta$ & 3.33 & 3.33 & 3.33 \\
$v_\star$ & 0.3 & 0.3 & 0.3 \\
$\alpha$ & $-1.27$ & $-1.27$ & $-1.27$ \\
$m_l$ & 5.736 & 5.736 & 5.736 \\
$N$ & 10,132 & 10,111 & 10,147 \\
\hline
\multicolumn{4}{l}{Inferred Parameters from Forward Model} \\
\hline
$\gamma$ & $10.502_{-0.004}^{+0.004}$ & $10.444_{-0.003}^{+0.004}$ & $10.450_{-0.005}^{+0.005}$ \\
$\beta$ & $3.355_{-0.031}^{+0.031}$ & $3.094_{-0.025}^{+0.024}$ & $3.164_{-0.037}^{+0.039}$ \\
$\sigma_m$ & $0.162_{-0.003}^{+0.003}$ & $0.145_{-0.003}^{+0.003}$ & $0.210_{-0.004}^{+0.004}$ \\
$v_\star$ & $0.264_{-0.033}^{+0.033}$ & $0.273_{-0.030}^{+0.033}$ & $0.248_{-0.037}^{+0.040}$ \\
$\alpha$ & $-1.206_{-0.040}^{+0.039}$ & $-1.225_{-0.043}^{+0.042}$ & $-1.181_{-0.043}^{+0.046}$ \\
\hline
\multicolumn{4}{l}{Inferred Parameters from Inverse Model} \\
\hline
$\gamma$ & $10.560_{-0.003}^{+0.004}$ & $10.500_{-0.003}^{+0.003}$ & $10.554_{-0.004}^{+0.005}$ \\
$\beta$ & $3.587_{-0.024}^{+0.025}$ & $3.337_{-0.024}^{+0.023}$ & $3.593_{-0.029}^{+0.029}$ \\
$\sigma_w$ & $0.046_{-0.001}^{+0.001}$ & $0.046_{-0.001}^{+0.001}$ & $0.062_{-0.001}^{+0.001}$ \\
\hline
\end{tabular}
\end{center}
\tablecomments{The fitted parameters are assumed to have flat priors within the following bounds: $10 < \gamma < 11$, $2.5 < \beta < 4.5$, $0.001 < \sigma_m < 0.3$, $0.001 < \sigma_w < 0.1$, $-1 < v_\star < 1$, and $-2 < \alpha < 0$. For both the forward and the dual-scatter models, the standard mass limit $m_l$ is fixed to the input value.}
\end{table}

To test whether the models in \S\,\ref{sec:likefun} can recover unbiased parameter estimation, I simulate three samples of galaxies where the ground truth of the model parameters are known. I start the process by random sampling of the velocity function in narrow redshift intervals. This is necessary because  other than the Schechter function parameters, the survey volume and the redshift distribution also change the probability of sampling a galaxy with a particular rotation velocity, and both of these factors could vary with redshift. So the first step is to divide the specified redshift range into a fine grid of equal intervals ($\delta cz$), then for each redshift, a subsample is generated following the procedure below:
\begin{enumerate}
\item The redshift grid center, $cz_0$, represents the true cosmological redshift of the simulated subset, i.e., all galaxies in this subset are at the same true distance $d(cz_0)$, calculated using Eq.\,\ref{eq:DL}. 
\item The initial subsample size for each $cz_0$ is calculated as $N_0 = a\, (cz_0)^{2+n}\, \delta cz$, where $a$ is a scale factor and $n$ is the power-law index of the integrated galaxy volume density. Both parameters can be adjusted to change the redshift distribution and the size of the final sample after the selection function is applied. 
\item {\it Optional}: Random redshifts perturbed by velocity noise, $\{cz_i\}_{i=1}^{N_0}$, are drawn around $cz_0$ following a normal distribution with a mean of $cz_0$ and a standard deviation of $\sigma_{cz}$: $\mathcal{N}(cz_0, \sigma_{cz}^2)$. Distances, $\{\tilde{d}_i\}$, are then calculated from $\{cz_i\}$ using Eq.\,\ref{eq:DL}. The velocity noise is thus propagated into $\{\tilde{d}_i\}$.
\item Intrinsic edge-on velocities, $\{v_i\}$, are drawn from the Schechter function in Eq.\,\ref{eq:VF} with specified parameters $(v_\star,\alpha)$ using the inverse transform sampling method. The default sampling range is $-3.5 < \beta v-v_\star < 1.5$.
\item Intrinsic masses, $\{M_i\}$, are calculated from the edge-on velocities using an idealized TFR: $M_i = \beta v_i + \gamma$. The exact linear relation between $\{v_i\}$ and $\{M_i\}$ makes sampling the intrinsic velocity function equivalent to sampling the intrinsic mass function in Eq.\,\ref{eq:MF}, as long as $M_\star$ is set to equal $v_\star + \gamma$.
\item Apparent masses, $\{\tilde{m}_i\}$, are calculated from the idealized TFR mass by subtracting the true distance and adding Gaussian scatters: $\tilde{m}_i = M_i - d(cz_0) + \epsilon_{m}$. The random scatter $\epsilon_m$ is drawn from a normal distribution with a standard deviation of $\sigma_m$: $\mathcal{N}(0,\sigma_m^2)$.
\item Projected velocities, $\{\tilde{w}_i\}$, are calculated from the edge-on velocities by adding random inclinations and Gaussian scatters: $\tilde{w}_i = v_i + \log \sin \epsilon_i + \epsilon_w$. To simulate samples with isotropic random orientations, the cosine of the inclination angle, $\cos \epsilon_i$, is drawn from a uniform distribution between 0 and 1, which then allow $\sin \epsilon_i$ to be computed as $\sqrt{1-\cos^2 \epsilon_i}$. The random scatter in $w$ is drawn from a normal distribution with a standard deviation of $\sigma_w$, i.e., $\epsilon_w \sim \mathcal{N}(0,\sigma_w^2)$. 
\item Finally, the observational selection function, $S(\tilde{m},\tilde{w})$, is applied to the simulated data set, and the survived subsample is kept.
\end{enumerate}
For simplicity, I use homoscedastic scatters in the simulated data (i.e., $\sigma_m$ and $\sigma_w$ are both constants for the entire sample), although the Bayesian methods can handle heteroscedastic measurement errors. It is worth noting that the frequently reported increase in mass/magnitude dispersion among slow rotators in a TF sample is mainly due to the smearing of the luminosity limit across the redshift range of the sample (see Fig.\,\ref{fig:data_dist} for example). Dispersion measurements in mass/magnitude near these luminosity limits should be deemed as upper limits, thus they do not offer evidence for heteroscedastic intrinsic dispersion in the TFR. 

Three simulated samples were generated for testing. They assume modest levels of scattering that are typically expected in galaxy samples like the ALFALFA or CF4:
\begin{enumerate}[label=\textbf{\Alph*.}]
\item $\sigma_{m}=0.15, \sigma_{w} = 0$ (no scatter in $w$)
\item $\sigma_{m}=0, \sigma_{w} = 0.045$ (no scatter in $m$)
\item $\sigma_{m}=0.15, \sigma_{w} = 0.045$ (scatters in both axes)
\end{enumerate}
Samples A and B match the assumptions of the forward model and the inverse model, respectively, and Sample C represents the more realistic case where scatters are present in both axes. The ratio of the dispersions, $\sigma_m/\sigma_w$, matches the assumed slope of the TFR ($\beta = 3.33$) to ensure equal contributions.

The other model parameters are shared across the samples:
\begin{itemize}
\item The velocity noise due to residual peculiar motion is ignored ($\sigma_{cz} = 0$) to avoid the additional marginalization over the true distance, as explained in \S\,\ref{sec:fTFR}. 
\item The selection function depends only on the apparent logarithmic mass ($\tilde{m}$) and is a step function with a detection limit at $m_l = 5.736$, appropriate for the ALFALFA sample.
\item The power-law index of the integrated galaxy volume density is set to $n = -1$ to produce a flatter redshift distribution, and the sampling scale factor is set to produce $\sim$10,000 galaxies over the redshift range between $4,000 < cz < 18,000$\,\kms\ above the detection limit. The sample size and its redshift distribution are comparable to those of \HI-selected galaxy samples like the ALFALFA and CF4. 
\item The TF parameters are set to $\beta = 3.33$ and $\gamma = 10.5$, similar to the values found in \citet{Kourkchi22}.
\item The Schechter function parameters for the intrinsic velocity and mass distributions are set to $\alpha = -1.27$, $v_\star = 0.3$, and $M_\star = v_\star + \gamma = 10.8$. These values are consistent with the tabulated baryonic mass function of \citet{Papastergis12}. 
\end{itemize}
The input parameters and the final sample sizes ($N$) are listed in Table\,\ref{tab:pars}. As an example, Fig.\,\ref{fig:data_dist} shows the distributions of simulated sample C in the plane of mass vs. projected velocity width and mass vs. redshift. The former illustrates the combination of Sine scatter and Gaussian scatters, while the latter illustrates the observational selection function.

\subsection{Code Validation and the General Eddington Bias} \label{sec:test_result}

\begin{figure}
\epsscale{0.95}
\plotone{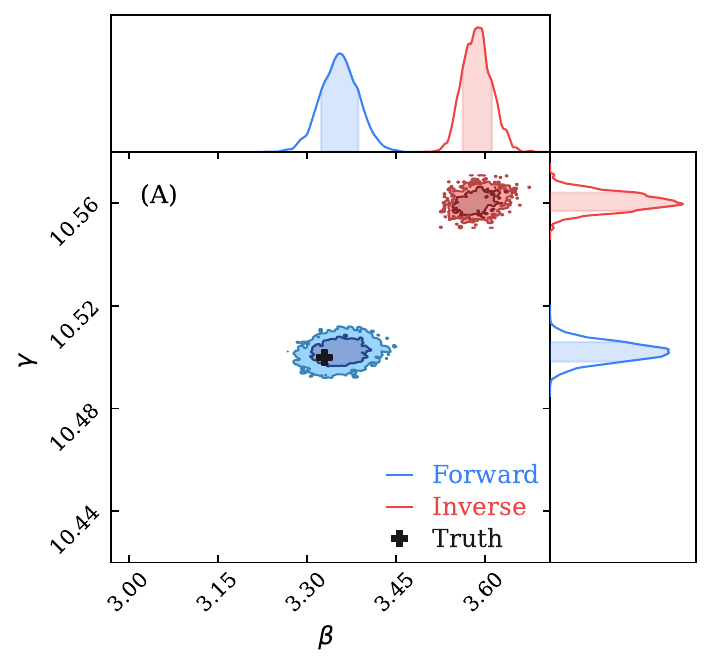}
\plotone{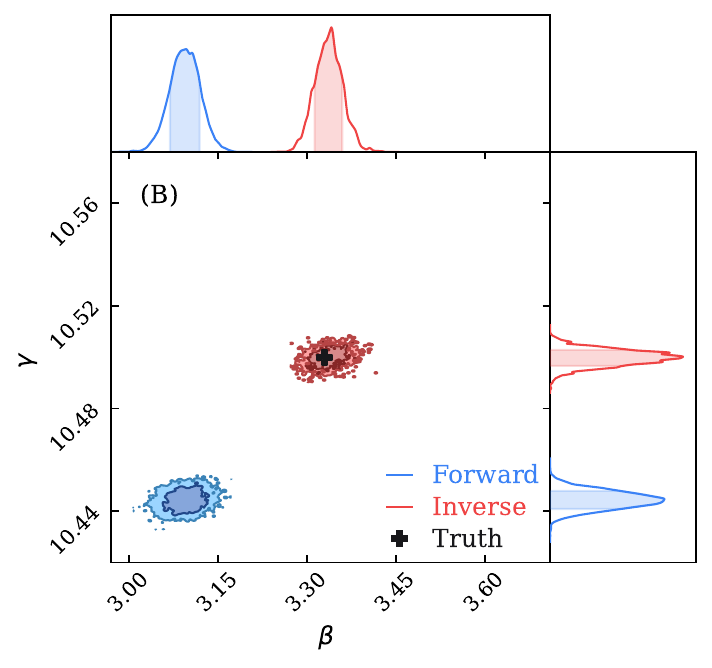}
\plotone{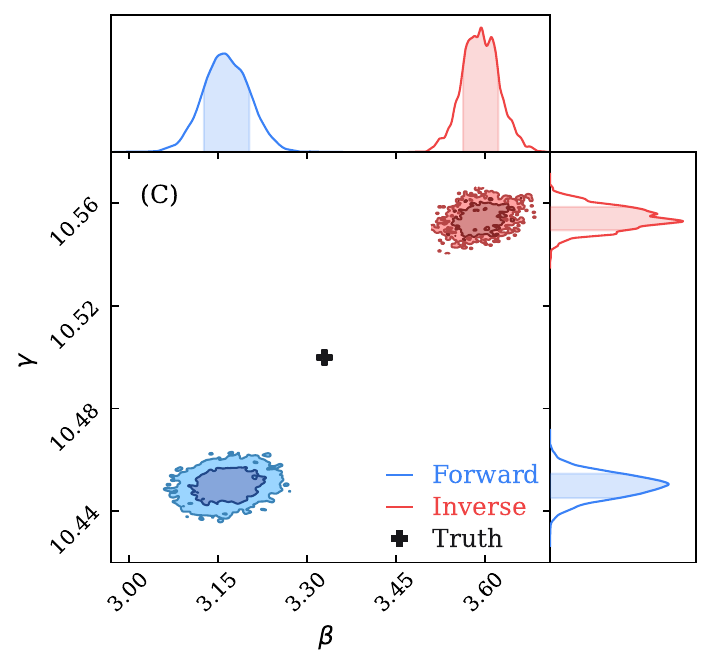}
\caption{MCMC-sampled posterior pdfs from the forward ({\it blue}) and the inverse ({\it red}) models for samples A, B, and C, respectively. For simplicity, only the two TF parameters are shown: slope ($\beta$) and intercept ($\gamma$). In all posterior plots, the contours in the joint pdfs enclose 68\% and 95\% of the volume, the highlighted regions in the marginalized pdfs show the credible intervals defined by the 16th and the 84th percentiles, and the black crosses indicate the truth. Biases appear whenever the data contains scattering in the model-assumed independent variable. 
\label{fig:fTFR_v_iTFR}} 
\epsscale{1.0}
\end{figure}

\begin{figure}
\epsscale{1.15}
\plotone{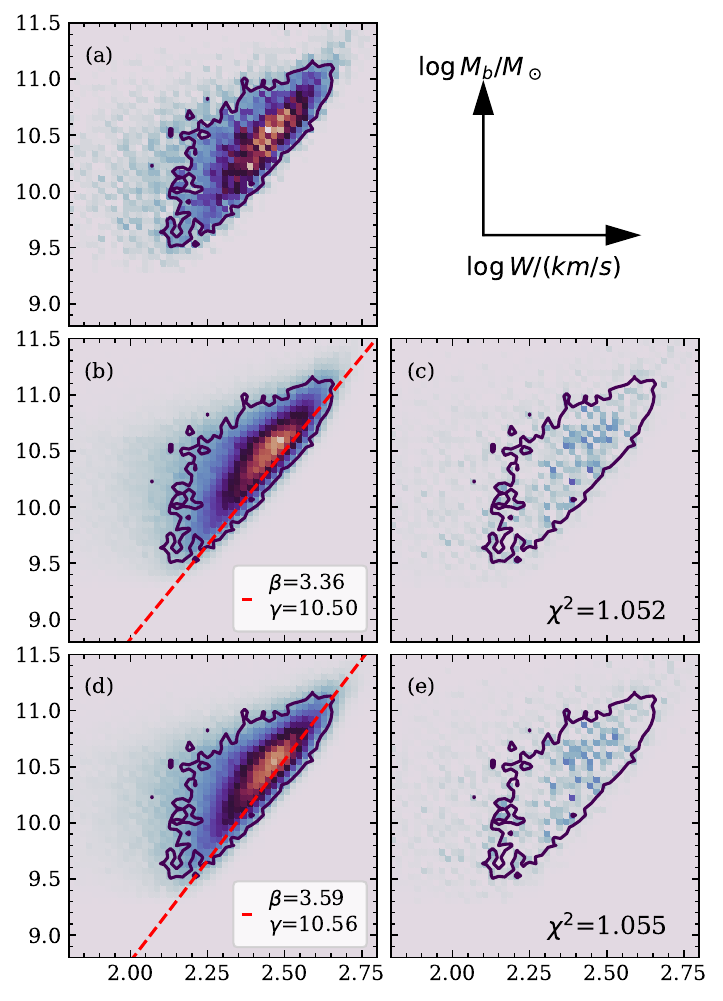}
\caption{The distribution of mass vs. projected line width of the input data ($a$) is compared with the distributions of larger samples simulated using the best-fit parameters of the forward model ($b$) and the inverse model ($d$). The contours enclose bins containing a minimum of 9 objects in panel $a$. In this example, the input data are from sample A, so only the forward model inferred the correct TF parameters. However, both models fit the distribution of the data equally well according to the residuals in panels $c$ and $e$. This result shows that it is difficult to use a goodness-of-fit parameter such as the reduced $\chi^2$ to select the correct model. 
\label{fig:fTFR_v_iTFR_data}} 
\epsscale{1.0}
\end{figure}

According to the theory of the general Eddington bias in \S\,\ref{sec:bias}, parameter estimation from unidirectional models would be unbiased only when there is no scatter in the independent variable; otherwise, the estimated parameters will be biased. The testing in this subsection thus serves two purposes: (1) to validate the implementation of the likelihood functions in the code, and more important, (2) to quantify the level of Eddington biases when there is scatter in the independent variable.

Assuming bounded flat priors for all parameters, I run the forward and the inverse models on the three simulated samples and the inferred parameters are compared to the input parameters in Table\,\ref{tab:pars}. The MCMC-sampled posterior pdfs of the two key parameters (TF slope $\beta$ and intercept $\gamma$) are presented in Fig.\,\ref{fig:fTFR_v_iTFR}. As expected, the forward model recovers the truth using Sample A, and the inverse model recovers the truth using Sample B. These findings provide confidence in the implemented likelihood functions of the two models. On the other hand, significant biases are observed when the forward model is used on Samples B and C (where $\sigma_w \ne 0$), and when the inverse model is used on Samples A and C (where $\sigma_m \ne 0$). 

The directions of the biases are that the forward model underestimate both parameters while the inverse model overestimate them. The amount of biases ($B_\gamma \simeq \pm0.06, B_\beta \simeq \pm0.24$) are significantly greater than the statistical uncertainties shown in the marginalized pdfs ($\sigma_\gamma \simeq 0.004, \sigma_\beta \simeq 0.03$). Further experiments show that both biases increase almost linearly with the variance of the independent variable (see \S\,\ref{sec:model_bias}).

It is worth noting that the same biases are present when inclination-corrected data are used; i.e., the bias is not introduced by treating inclination as a latent variable. Instead, it is a feature of unidirectional regression models. As mentioned earlier, it is well known that the OLS estimate of the regression slope is biased to zero when the independent variable is measured with error \citep[e.g.,][]{Fuller87, Akritas96}. Since least-squares is a maximum likelihood estimator, the same biases are expected in unidirectional MLE methods that neglect the error of the independent variable, regardless whether the data are corrected for inclination or not.

Despite of the biased parameter estimation, both models fit the input data equally well for all three input samples. I demonstrate this with sample A in Fig.\,\ref{fig:fTFR_v_iTFR_data}, where simulated samples 20$\times$ larger than the input sample are generated using the best-fit parameters of the two models\footnote{Unconstrained $\sigma$ is set to zero and unconstrained Schechter function parameters are set to the truth values.}, and their distributions are compared with that of the input sample. The model residuals show statistical fluctuations comparable to the Poisson noise, as indicated by the reduced $\chi^2$ values around unity. The slight difference in the $\chi^2$ values between the two models is actually due to the stochastic nature of the simulated samples. This finding shows that the goodness-of-fit cannot be used to tell which model is better supported by the data. Instead of model selection, one should focus on methods that can mitigate the general Eddington bias, which are the topics of the next section.   

\section{Unbiased Parameter Inference} \label{sec:debias}

This section introduces two latent-inclination methods that can mitigate both Malmquist bias and Eddington bias in the inference of the TFR. The first method shifts the dependent variable using the predicted Eddington bias from its analytical expression (\S\,\ref{sec:iTFR_Edd}), and the second method incorporates the scatter of the independent variable in the likelihood function of a bidirectional dual-scatter model (\S\,\ref{sec:uTFR}). Technically, there is a third method---choosing an empirical unbiased anchor point for the correlation (\S\,\ref{sec:model_bias})---but it only reduces the bias of the intercept parameter, limiting its application. 

\subsection{Shifting the Moment of the Dependent Variable} \label{sec:iTFR_Edd}

\begin{figure}
\epsscale{1.15}
\plotone{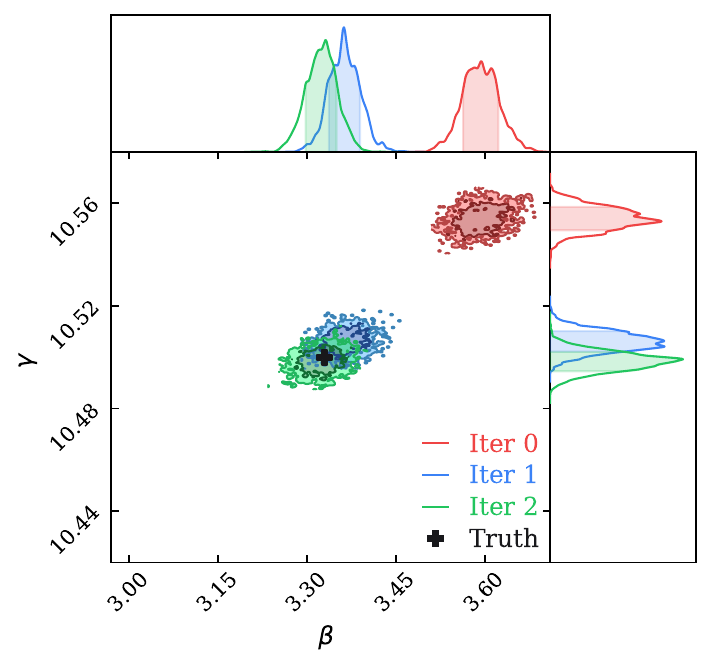}
\caption{MCMC-sampled posterior pdfs from the inverse model after iterative corrections of the general Eddington bias. Sample C is used as the input data. Iteration 0 is before any bias correction, so it is the same as the posterior of the inverse model in Fig.\,\ref{fig:fTFR_v_iTFR}$c$. In subsequent iterations, $\{\tilde{w}_k\}_{k=1}^N$ are bias corrected using Eq.\,\ref{eq:iTFR_Edd_Corr}. The process converges after just a few iterations, but unbiased parameter inference is achieved only when the correct $\sigma_m$ is specified. 
\label{fig:iTFR_Edd_Corr}} 
\epsscale{1.0}
\end{figure}

In \S\,\ref{sec:bias}, I have explained that the general Eddington bias is caused by the scatter in the independent variable and the gradient of its distribution function. It is thus expected that, when the predicted Eddington bias of the dependent variable is corrected for using its analytical expression, the inferred regression coefficients will be no longer biased.

For calibrating the TFR using inclination-corrected data, one could choose to debias the dependent variable in either the forward model or the inverse model. But when the inclination angle is considered a latent variable, one can only correct $\tilde{w}$ for the inverse model. This is because to debias $\tilde{m}$ for the forward model using Eq.\,\ref{eq:Edd_gen_y} requires knowing the inclination-corrected velocity width ($x \equiv \tilde{w}-i$). 

For the forward model, the general Eddington bias of the mean $y$ at a fixed $\tilde{x}$ is given by Eq.\,\ref{eq:Edd_gen_y} in \S\,\ref{sec:bias}:
\begin{equation}
y(\tilde{x}) - \langle y \rangle_{\tilde{x}} = -\beta\sigma_x^2 \frac{d \ln p(\tilde{x})}{d\tilde{x}} 
\end{equation}
For the inverse model, the general Eddington bias of the mean $x$ at a fixed $\tilde{y}$ is:
\begin{equation} \label{eq:Edd_gen_x}
x(\tilde{y})-\langle x \rangle_{\tilde y} = -\frac{\sigma_y^2}{\beta} \frac{d \ln p(\tilde{y})}{d \tilde{y}}
\end{equation} 
To remove the bias, one can shift the value of individual $x$:
\begin{equation}
x_c = x - \frac{\sigma_y^2}{\beta} \frac{d \ln p(\tilde{y})}{d \tilde{y}}
\end{equation}
so that the mean of $x_c$ matches the $\tilde{y}$-predicted $x$:
\begin{equation}
x(\tilde{y}) - \langle x_c \rangle_{\tilde{y}} = x(\tilde{y})-\langle x \rangle_{\tilde{y}} + \frac{\sigma_y^2}{\beta} \frac{d \ln p(\tilde{y})}{d \tilde{y}} = 0
\end{equation}

If $\tilde{y}$ follows a Schechter function:
\begin{equation}
p(\tilde{y}) \propto 10^{(\alpha+1)(\tilde{y}-y_\star)} \exp(-10^{\tilde{y}-y_\star})
\end{equation}
then the Eddington bias correction in $x$ is:
\begin{equation} 
x_c = x - \frac{\ln 10}{\beta} \sigma_y^2 [(\alpha + 1) - 10^{\tilde{y}-y_\star}]
\end{equation}
For the case of the inverse TFR, the bias correction in $\tilde{w}$ is:
\begin{equation} \label{eq:iTFR_Edd_Corr}
\tilde{w}_c = \tilde{w} - \frac{\ln 10}{\beta} \sigma_m^2 [(\alpha + 1) - 10^{(\tilde{m}+d)-M_\star}]
\end{equation}
which shows that the correction depends on four parameters $(\sigma_m, \beta, M_\star, \alpha)$ and the observed mass of the galaxy ($\tilde{m}+d$). The success of the bias correction depends on  the accurate specification of the four parameters.

First, $M_\star$ and $\alpha$ characterize the observed mass function, which can be estimated through various methods. For example, it can be measured using luminosity-function estimators that correct for the Malmquist bias, e.g., the $C^-$ method \citep{Lynden-Bell71}, followed by a Schechter function fit. Here I opt to infer $v_\star$ and $\alpha$ using the latent-inclination forward model introduced in \S\,\ref{sec:fTFR}, and calculate $M_\star$ as $v_\star + \gamma$. It is safe to use $v_\star$ and $\alpha$ from the forward model because (1) they are mildly biased (unlike $\beta, \gamma$), i.e., their deviations from the truth are within twice the statistical error (see the blue curves in the third and fourth columns of Fig.\,\ref{fig:uTFR}), and (2) the Eddington bias correction is relatively insensitive to these parameters.

Next, $\beta$ is the slope of the TFR, which is a key parameter to be determined by the model. Using it as an input parameter thus requires an iterative process. It starts from the results of the forward model and the inverse model without Eddington bias correction. The former provides $v_\star$, $\alpha$, and the upper limit on $\sigma_m$, the latter provides $\beta_j$ and $\gamma_j$ for $j = 0$, where the subscripts indicate that they are the initial guess values and will be updated in subsequent iterations. For a specified value of $\sigma_m$, the line widths are then corrected for the Eddington bias using Eq.\,\ref{eq:iTFR_Edd_Corr} with parameters $(\sigma_m, \beta_j, v_\star+\gamma_j, \alpha)$. The subsequent runs of the inverse model uses Eddington-bias-corrected line widths ($\tilde{w}_c$) as the input data and produce updated values of $\beta_j$ and $\gamma_j$. Fig.\,\ref{fig:iTFR_Edd_Corr} illustrates the iterative process and shows that results converge in just a few iterations. 

Finally, $\sigma_m$ is the mass scatter that includes both measurement error and intrinsic dispersion. This is the only parameter that needs to be specified by the user based on prior knowledge of the data. In the worse case, one can use the forward model to estimate the total dispersion of the data in the $m$-direction ($\sigma_m^t$), which gives the upper limit of $\sigma_m$: $0 < \sigma_m < \sigma_m^t$. The bias-corrected results should lie somewhere between the two extremes defined by the forward and the inverse models without corrections (e.g., see Fig.\,\ref{fig:fTFR_v_iTFR} for Sample C). Underestimating $\sigma_m$ causes under-correction of the Eddington bias, so the inferred $\beta$ and $\gamma$ will be biased high (although their biases are reduced compared to those from the uncorrected model). On the other hand, overestimating $\sigma_m$ causes over-corrections of the Eddington bias, so the TF parameters overshoot the truth and will be biased low as they approaches those inferred from the forward model. Therefore, the primary limitation of this method is that the inferred parameters will be unbiased only when the correct value of $\sigma_m$ is prescribed by the user. 

\subsection{The Bidirectional Dual-Scatter Model} \label{sec:uTFR}

\begin{figure*}
\epsscale{1.15}
\plotone{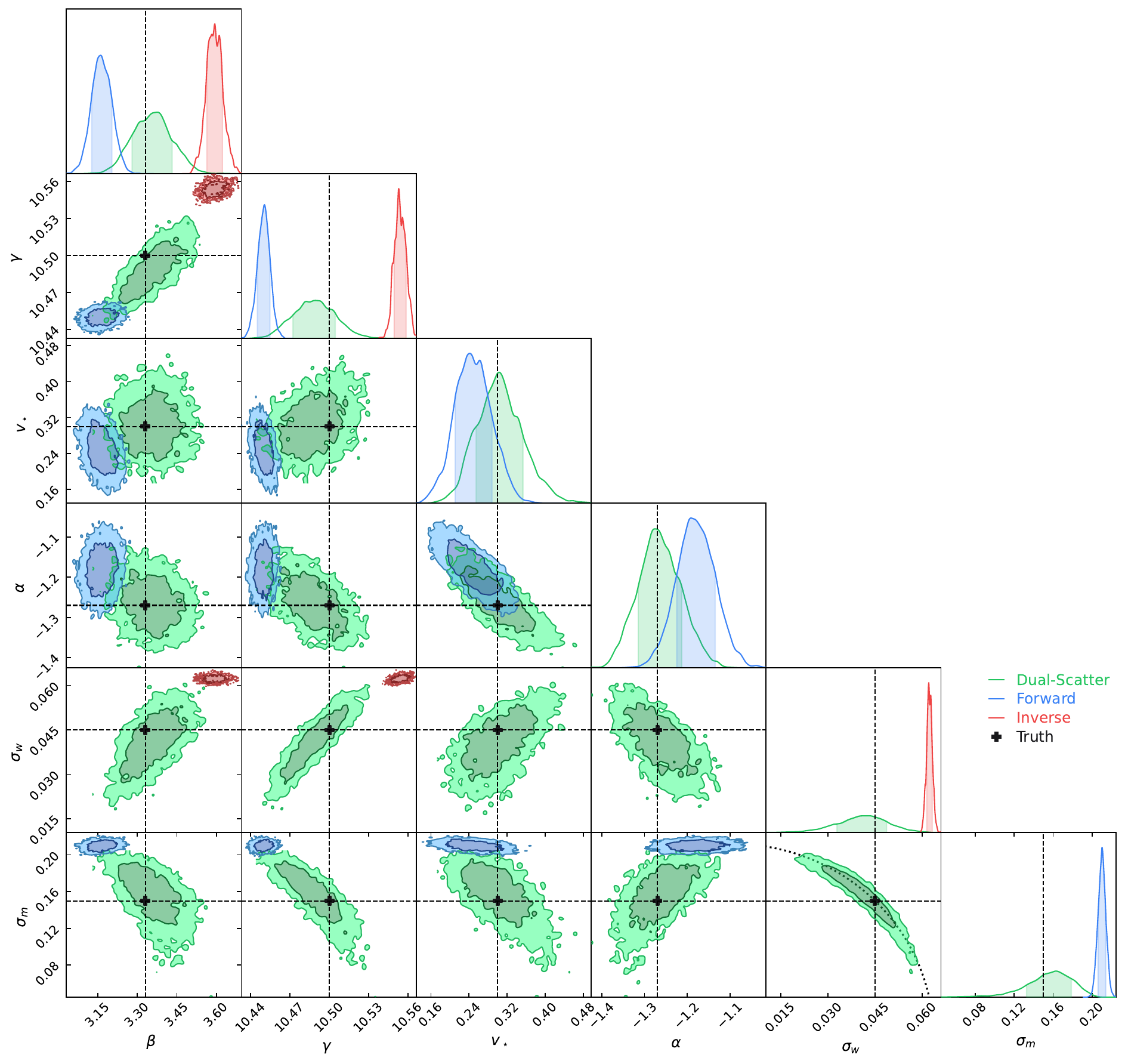}
\caption{MCMC-sampled posterior pdfs from the dual-scatter model ({\it green}) compared to those from the forward model ({\it blue}) and the inverse model ({\it red}). The data are from sample C, whose input parameters are indicated by black dashed lines and black crosses. The dual-scatter model produces relatively unbiased inferences for all parameters, gives more realistic statistical uncertainties, and reveals parameter degeneracies. For example, the two $\sigma$'s are tightly correlated on an ellipse: $\sigma_m^2 + \beta^2 \sigma_w^2 = \sigma_t^2$, as indicated by the dotted curve in the panel.
\label{fig:uTFR}} 
\epsscale{1.0}
\end{figure*}
 
In this subsection, I introduce the dual-scatter model where I expand the likelihood function of the latent-inclination forward model by including scatters of the independent variable $w$. In Appendix\,\ref{sec:unify_iTFR}, I expand the likelihood function of the inverse model by including scatters in $m$. It is shown there that the results are mathematically equivalent, making the model bidirectional (symmetric). 

When $\tilde{w}$ and $\tilde{m}$ are independent\footnote{This is a valid assumption because correlated measurement errors in TFR data are introduced by inclination correction.}, the joint probability of the dual-scatter model can be decomposed as:
\begin{align} \label{eq:joint_uTFR}
p(\tilde{m},\tilde{w},d,w,i) &= p(\tilde{m},\tilde{w},d | w,i) p(w,i) \nonumber \\
&= p(\tilde{m}|d,w,i) p(\tilde{w}|w) p(d) \times p(w|i) p(i) \nonumber \\
&= p(\tilde{w}|w) p(\tilde{m},w,d,i)
\end{align} 
where $p(\tilde{m},w,d,i)$ is given by Eq.\,\ref{eq:joint_fTFR2}, and the new term, $p(\tilde{w}|w)$, is a Gaussian:
\begin{equation}
p(\tilde{w}|w) = \frac{1}{\sigma_w\sqrt{2\pi}} \exp \left[ -\frac{(\tilde{w}-w)^2}{2\sigma_w^2} \right]
\end{equation}

The joint pdf of the observables $(\tilde{m},\tilde{w},d)$ is then calculated by marginalizing Eq.\,\ref{eq:joint_uTFR} over both $w$ and $i$:
\begin{align} \label{eq:joint_uTFR2}
&p(\tilde{m},\tilde{w},d) \nonumber \\
&= \int_{-\infty}^{\infty} \int_{-\infty}^0 p(\tilde{m},\tilde{w},d,w,i)\,di\,dw \nonumber \\
&= \frac{\beta \phi_\star p(d)}{2 \pi \sigma_m \sigma_w} 
\int_{-\infty}^{\infty} \int_{-\infty}^0 \frac{\ln 10 \cdot 10^{2 i }}{\sqrt{1-10^{2 i}}} \nonumber \\
&\times \exp \left[-\frac{(\tilde{w}-w)^2}{2\sigma_w^2}\right] 
\exp \left[-\frac{(\tilde{m}+d-\beta(w-i)-\gamma)^2}{2\sigma_m^2}\right]  \nonumber \\
&\times 10^{(\alpha+1)[\beta(w-i)-v_\star]} \exp (-10^{\beta(w-i)-v_\star})\,di\,dw
\end{align}

Accounting for the selection function $S(\tilde{m},\tilde{w})$, the conditional pdf of $\tilde{m}$ is: 
\begin{equation} \label{eq:cpdf_uTFR}
p(\tilde{m} | \tilde{w}, d) = 
\frac{S(\tilde{m},\tilde{w}) p(\tilde{m}, \tilde{w}, d)}{\int_{-\infty}^{\infty} S(\tilde{m},\tilde{w}) p(\tilde{m}, \tilde{w}, d) \, d\tilde{m}} 
\end{equation}
And the data likelihood function is composed of the above conditional pdfs:
\begin{equation} \label{eq:lnL_uTFR}
\ln {\mathcal L} = \sum_{k=1}^N \ln p(\tilde{m}_k | \tilde{w}_k, d_k)
\end{equation} 

Like the forward model, when the selection function is a step function of $\tilde{m}$, the integral over $\tilde{m}$ on the denominator of Eq.\,\ref{eq:cpdf_uTFR} becomes an error function. This replacement reduces the triple integral to a double integral. I provide the full expression of the conditional pdf in Eq.\,\ref{eq:cpdf_uTFR_step} in Appendix\,\ref{sec:step_fun}. 

Note that even after this simplification, evaluating the likelihood function remains computationally expensive because each data point still requires calculating two double integrals. In Appendix\,\ref{sec:accelerate}, I describe a Fast Fourier Transform (FFT) method that drastically accelerated the computation of the double integrals. Additional acceleration is achieved by vectorizing the conditional pdf and leveraging the integrated Graphics Processing Unit (GPU). Together, the computation time is reduced by three orders of magnitude when compared to direct integration, making the dual-scatter model even faster than the forward model, when the latter is run on the CPU.

The dual-scatter model is applied to the simulated Sample C, which includes significant amounts of scatter in both axes. I show the MCMC-sampled posterior pdfs in Fig.\,\ref{fig:uTFR} and compare them with those from the forward and the inverse models. The results from the dual-scatter model can be summarized as follows:
\begin{itemize}
\item The marginalized posteriors show that the best-fit (median) parameters are relatively unbiased for all three samples, i.e., they lie within $1\sigma$ from the truth. 
\item The statistical uncertainties of the inferred parameters are larger compared to those from the forward and the inverse model: the TF parameters $(\beta, \gamma)$ show 2-3$\times$ larger uncertainties, the dispersion parameters $(\sigma_m,\sigma_w)$ show 2-8$\times$ greater uncertainties, while the Schechter parameters $(v_\star, \alpha)$ show comparable uncertainties.
\item The joint posteriors show significant correlations among parameters. In particular, the two dispersion parameters are tightly correlated on an ellipse: $\sigma_m^2 + \beta^2 \sigma_w^2 = \sigma_t^2$. These covariances explain the larger uncertainties of the inferred parameters.
\end{itemize}

Compared to the forward model, including the scatter of the independent variable $w$ requires multiplication of another Gaussian function of $\tilde{w}-w$ and an additional marginalization over $w$ in both the numerator and the denominator of the conditional pdf that constitutes the likelihood function. The added complexity makes the MCMC sampling significantly more computationally expensive, motivating innovative applications of numerical methods and GPU acceleration. The resulting posteriors show nearly unbiased parameter estimation, more realistic estimates of the uncertainties, and strong degeneracy between the two dispersion parameters. The degeneracy is a feature of the mathematical problem, so it cannot be eliminated, but its effects decreases as the sample size increases.  

\subsection{The Unbiased Anchor Point} \label{sec:model_bias}

The TFR is typically anchored at $\log V_0 = 2.5$, which is an arbitrary choice. In this subsection, I describe the existence of a preferred anchor point at which the inferred intercepts of both the forward and the inverse models are unbiased. 

By simulating a number of data sets that share the same TFR ($\beta = 3.33, \gamma = 10.5$) and observational selection function but have different amount of scatters in the independent variable ($w$), I find the following power-law relations between the biases of the regression coefficients and the amount of scatter ($\sigma_w$):   
\begin{align} \label{eq:model_bias}
B_\gamma =
\begin{cases}
-0.058 (\sigma_w/0.05)^{1.8} & \text{forward model} \\
+0.060 (\sigma_m/0.15)^{1.8} & \text{inverse model} 
\end{cases} \nonumber \\
B_\beta =
\begin{cases}
-0.246 (\sigma_w/0.05)^{1.8} & \text{forward model} \\
+0.237 (\sigma_m/0.15)^{1.8} & \text{inverse model}
\end{cases}
\end{align}

Notice that because the biases of both coefficients scale with $\sigma_w$ following power laws of the same index, their ratios are independent of the scatter. A consequence of the constant bias ratio $B_\gamma/B_\beta$ is that the true TFR and the inferred TFRs from the two models all converge at a single point at:
\begin{align} \label{eq:unbiased_anchor}
\log V_0 &= 2.5 - \frac{B_\gamma}{B_\beta} \nonumber \\
\gamma_0 &= \gamma - \beta \frac{B_\gamma}{B_\beta} 
\end{align}
At this converging point, the intercepts of all the inferred TFRs are unbiased. This finding suggests that one can reduce the general Eddington bias by choosing the unbiased anchor location instead of the default anchor at $\log V_0 = 2.5$.

This unbiased anchor location depends on sample characteristics and the coefficients of the TFR, but it can be determined empirically by running the forward model and the inverse model on the same data. Take Sample C as an example, I use the inferred TF parameters listed in the last column of Table\,\ref{tab:pars}. I first calculate the differences between the inferred parameters of the two models, $\Delta B_\gamma = 10.554-10.450 = 0.104, \Delta B_\beta = 3.593-3.164 = 0.429$, then their ratio, $\Delta B_\gamma/\Delta B_\beta = 0.24$, and finally the unbiased anchor point at $\log V_0 = 2.5 - \Delta B_\gamma/\Delta B_\beta = 2.26$. At this anchor point, the intercepts of both the forward model and the inverse model are at $\gamma_0 = 9.69$, which is essentially the same as the intercept of the true relation at $\log V_0 = 2.26$ ($\gamma_0 = 9.70$). Despite of their different slopes, adopting the converging point as the anchor has minimized the bias of the inferred intercepts for both the forward and the inverse models.

\section{Summary and Discussion} \label{sec:summary}

Linear regression of the TFR requires the observed data to be corrected for inclination. But the current method to estimate inclination angle from galaxy morphology is highly uncertain and requires well-resolved images. One solution is to treat the inclination as a latent variable with a known distribution function, so that one can infer the TFR using data uncorrected for inclinations. In this paper, I have developed Bayesian latent-inclination inference methods that can also mitigate the distance-dependent Malmquist bias due to sample selection and the general Eddington bias due to scatter in the independent variable. These biases are important because they directly impact measurements of the Hubble constant.

I start by deriving the likelihood functions for two latent-inclination unidirectional models that neglect the scatter of the independent variable: the forward model (velocity width as independent variable) and the inverse model (mass as independent variable). Both likelihood functions are implemented in Python with a MCMC-sampler and are tested on simulated galaxy samples with random orientations and with random Gaussian scatters in mass and velocity. I find that, when the independent variable has significant scatters, the inferred regression coefficients are biased (Fig.\,\ref{fig:fTFR_v_iTFR}), although biased models demonstrate similarly good quality of fit to the data (Fig.\,\ref{fig:fTFR_v_iTFR_data}). The biases of both coefficients increase with the scatter of the independent variable $B_\theta \propto \sigma_x^{1.8}$. For example, using the inverse TFR model, neglecting a scatter of 0.15\,dex in luminosity would lead to an overestimate of $H_0$ by $\sim$5\,\hunit\ (Eq.\,\ref{eq:H0_bias}), comparable to the ``Hubble tension''.

Evidently, the scatter of the independent variable ($x$) does not simply propagate into the dependent variable ($y$) through the correlation as $\sigma_y = \beta \sigma_x$, because that would not have led to biases in the regression coefficients. Instead, the bias shifts the first moment of $y$ at each $x$ by altering the conditional pdf $p(y|x)$ according to Bayes' rule. I derive the analytical form of this bias (Eq.\,\ref{eq:Edd_gen_y}). Because the classic Eddington bias is a special case of this bias, it is termed the ``general Eddington bias''. 

Methods aimed at unbiased parameter inference must include both (1) the scatter of the dependent variable and the selection function to account for the distance-dependent Malmquist bias, and (2) the scatter and the distribution function of the independent variable to account for the general Eddington bias. I then introduce two approaches to mitigate the general Eddington bias: 
\begin{itemize}
\item Shifting the dependent variable by reversing the expected amount of Eddington bias ({\it Moment Shifting}, \S\,\ref{sec:iTFR_Edd}),
\item Incorporating the scatter of the independent variable in the likelihood function ({\it Dual-scatter Model}, \S\,\ref{sec:uTFR}).
\end{itemize}

Testing on simulated data sets shows that both methods can effectively reduce or eliminate both the Malmquist bias and the Eddington bias (Figs.\,\ref{fig:iTFR_Edd_Corr} and \ref{fig:uTFR}). Each method has its own strengths and limitations, and it is a trade-off between efficiency and accuracy: 
\begin{itemize}
\item The moment-shifting method can iteratively determine most of the required parameters for bias correction, but the dispersion parameter of the independent variable must be specified {\it a priori}. 
\item The dual-scatter model is the most versatile and is recommended for most scenarios. It fits both dispersion parameters simultaneously, produces more realistic estimates of parameter uncertainties, and reveals covariances among model parameters. But it is the most computationally expensive method, requiring FFT techniques and GPUs to accelerate (Appendix\,\ref{sec:accelerate}).
\end{itemize}

Technically, there is one additional method to mitigate the general Eddington bias: defining the intercept of the TFR at an unbiased anchor point (\S\,\ref{sec:model_bias}). The ``unbiased anchor'' method takes advantage of the correlation between the biases of the regression coefficients and determines the unbiased anchor empirically. The inferred intercepts at this preferred anchor are unbiased for both unidirectional models. However, the inferred slopes remain biased, and the location of the unbiased anchor depends on sample characteristics. This variability complicates applying the method to Hubble constant measurements, as the unbiased anchors for the redshift and zero-point samples are likely distinct.

Naturally, the next step is to apply the latent-inclination likelihood-based methods on actual data. Currently there are two major data sets for TF studies in the nearby Universe ($cz < 20,000$\,\kms): the ALFALFA sample of $\sim$31,500 galaxies\footnote{The ALFALFA sample contains 25,432 high S/N \HI\ detections (Code 1) and 6,068 low S/N detections with prior optical detection (Code 2).} \citep{Haynes18, Durbala20} and the CF4 BTF-Distances catalog of $\sim$10,000 galaxies\footnote{The CF4 catalogs are available in the Extragalactic Distance Database: \url{https://edd.ifa.hawaii.edu}.}  \citep{Kourkchi22}. The ALFALFA sample is better suited for the latent-inclination methods because it is a 7000\,deg$^2$ \HI\ survey from a single instrument (Arecibo ALFA) and it includes galaxies at all inclinations. In contrast, the CF4 sample is a compilation of \HI\ surveys from seven radio telescopes and excludes galaxies with estimated inclination less than 45$^\circ$. Some of the compiled subsamples in CF4 were selected to be edge-on systems, causing a spike near 90$^\circ$ in the distribution of measured inclinations.
The ALFALFA sample is also preferred because of selection bias: a single observational selection function that describes the full data set can be used in the likelihood function. For a heterogenous sample like the CF4, the complex selection effects from various instruments are unlikely to be captured by a single post facto selection function. Instead, each constituent subsample should be separately analyzed using their own selection function, and the final result would be a weighted average of the results from the subsamples.

Future applications of the methods on the ALFALFA data also face challenges. First, the \HI\ flux limit increases with line width. Like all spectroscopic line surveys with fixed integration time per area, ALFALFA is more sensitive to narrower lines because the same amount of signal is spread over fewer spectroscopic channels. The line flux limits set gas mass limits, making the latter also depending on the line width. Second, both the total \HI\ flux and the optical photometry become unreliable for the nearest galaxies. The former has been addressed by the special catalog for extended sources \citep{Hoffman19}. But to address the latter, one needs to replace the automatic photometry from pipelines with more involved asymptotic photometry \citep{Courtois11} for a large sample of galaxies. Because of the inclination cutoff at 45$^\circ$, only 43\% (13,617) of the ALFALFA sample are included in the CF4 Initial Candidates catalog, leaving 57\% of the sample without asymptotic photometry. 

Thus far, the discussion has centered on the linear baryonic TFR. Depending on the available data, researchers may prefer a different form of the TFR. Fortunately, the expressions derived for the baryonic TFR can be readily adapted to other empirical forms of the TFR:
\begin{itemize}
\item To apply the methods to magnitude-based TFR, only a simple modification is required: compute $m$ as $-0.4$ times the apparent magnitude at wavelength $\lambda$ (i.e., $m \equiv -0.4 m_\lambda$) to convert the decreasing magnitude scale defined by Pogson's ratio to an increasing decade scale. With this conversion, the best-fit intercept parameter $\gamma$ is related to the fiducial absolute magnitude $M_\lambda^0$ as: $M_\lambda^0 = -2.5\gamma - 25$. 
\item To apply the methods to non-linear TFR (e.g., to capture the curvature of the $i$-band TFR at the luminous end), one can replace the linear relation in Eq.\,\ref{eq:TFRshort} with more complex functions of $(w-i)$ (e.g., a polynomial).
\end{itemize}

I have made the following simplifications to keep the discussion focused on the Sine scatter of the velocity width and the Gaussian scatters of logarithmic mass and velocity width: (1) the redshifts are assumed to have been corrected for peculiar velocities, (2) the residual velocity noise is ignored, and (3) the apparent baryonic mass is assumed to be unaffected by galaxy inclination (i.e., either gas mass dominates or stellar mass extinction is accounted for by the color-derived mass-to-light ratio). Below I describe future extensions of the model to relax these assumptions.

Incorporating velocity noise in the likelihood function requires multiplying a Gaussian function of $\tilde{d}-d$ to the joint pdf and an additional marginalization over the true distance $d$ (\S\,\ref{sec:fTFR}). In addition, models that simultaneously fit the scale parameters of a peculiar velocity model and the velocity noise have been developed previously for TFR studies using inclination-corrected data \citep[e.g.,][]{Willick97,Boubel24a}. The same methodology can be applied to the latent-inclination models. 

For magnitude-based TFRs, to account for the inclination-dependent internal extinction in the forward and the dual-scatter model, one needs to replace $\tilde{m}$ with its extinction-corrected counterpart $\tilde{m}-0.4 A({\rm inc})$ in $p(\tilde{m}|w,i,d)$ (Eq.\,\ref{eq:gaus_m}). A common way to parameterize the extinction is $A({\rm inc}) = \tau \log [\sec ({\rm inc})]$ \citep[e.g.,][]{Shao07}. The extinction-corrected mass is then $\tilde{m}-0.2 \tau \log (1-10^{2i})$ (where $i \equiv \log [\sin ({\rm inc})]$ as defined in Eq.\,\ref{eq:shorthands}), and the conditional pdf becomes:
\begin{align}
&p(\tilde{m}|w,i,d) = \frac{1}{{\sigma_m \sqrt {2\pi } }} \nonumber \\
&\times \exp \left[ -\frac{[\tilde{m}-0.2 \tau \log(1-10^{2i})+d-\beta(w-i)-\gamma]^2 }{2\sigma_m^2 } \right]
\end{align}
The replacement adds $\tau$ (the face-on extinction) as a new parameter and changes the subsequent calculation of the conditional pdf $p(\tilde{m}|w,d)$ in Eq.\,\ref{eq:cpdf_fTFR} and $p(\tilde{m}|\tilde{w},d)$ in Eq.\,\ref{eq:cpdf_uTFR}.

To account for the inclination-dependent extinction in the inverse model, one needs to replace $m$ with $m-0.2 \tau \log (1-10^{2i})$ in both $p(\tilde{w}|m,d,i)$ and $p(m|d)$ (Eqs.\,\ref{eq:gaus_w} and \ref{eq:pmd_iTFR}). This modification causes the mass function term, $p(m|d)$, to depend on inclination, preventing it from being cancelled out in the conditional pdf in Eq.\,\ref{eq:cpdf_iTFR}.  

Latent-variable Bayesian methods provide wide applicability in observational astronomy. Although this paper focuses on inferring the Tully-Fisher relation, the likelihood-based framework is highly generalizable and can be adapted to other linear regression problems involving Gaussian and non-Gaussian scatter. The primary requirement is knowledge of the latent variable's prior probability distribution, which can be expressed analytically or provided numerically as a tabulated function.
 
\begin{acknowledgments} 
I appreciate helpful discussions with my colleagues Ken Gayley, Kevin Hall, and Steve Spangler. This work is supported by the University of Iowa through the Investment in Strategic Priorities (ISP) program and the National Science Foundation (NSF) grant AST-2103251. 
\end{acknowledgments}

\vspace{5mm}
\software {
	NumPy \citep{Harris20},
	SciPy \citep{Virtanen20}, 
	emcee \citep{Foreman-Mackey13},
	PyTorch \citep{Paszke19},
	ChainConsumer \citep{Hinton16}
	}

\bibliographystyle{apj}
\bibliography{TFR_biases}

\appendix

\section{The Conditional Probabilities of Flux-Limited Samples} \label{sec:step_fun}

In this Appendix, I present the complete analytical expressions for the conditional PDFs of the forward, inverse, and dual-scatter models for a common scenario where the sample is flux-limited and the selection function is a step function:
\begin{equation} \label{eq:step}
    S(\tilde{m})= 
\begin{cases}
    1,              & \text{if } \tilde{m} \geq m_l\\
    0,              & \text{otherwise}
\end{cases}
\end{equation}

This particular selection function significantly simplifies the marginalization of the joint pdf over $\tilde{m}$, because the only $\tilde{m}$-dependent term is a Gaussian, and its integral is the complementary error function (erfc):
\begin{align}
\int_{-\infty}^{\infty} S(\tilde{m}) p(\tilde{m}|w,i,d) \, d\tilde{m}
&= \int_{m_l}^{\infty} p(\tilde{m}|w,i,d) d\tilde{m} \nonumber \\ 
&= \int_{m_l}^{\infty} \frac{1}{{\sigma_m \sqrt {2\pi } }} \exp \left[ -\frac{[\tilde{m}+d-\beta(w-i)-\gamma)]^2 }{2\sigma_m^2 } \right] d\tilde{m} \nonumber \\
&= \frac{1}{2} {\rm erfc}\left[\frac{m_l+d-\beta(w-i)-\gamma}{\sqrt{2}\sigma_m}\right]
\end{align}

After this simplification, the conditional pdfs for the forward and the dual-scatter models become:
\begin{itemize} 
\item The forward model:
\begin{equation} \label{eq:cpdf_fTFR_step}
p(\tilde{m} | w, d) = 
\frac{\sqrt{2}}{\sqrt{\pi} \sigma_m} 
\frac{\int_{-\infty}^0 
\frac{10^{2i}}{\sqrt{1-10^{2i}}} 
\exp \left[-\frac{[\tilde{m}+d-\beta(w-i)-\gamma]^2}{2\sigma_m^2}\right] 
10^{(\alpha+1)[\beta(w-i)-v_\star]} \exp (-10^{\beta(w-i)-v_\star})  \, di}
{\int_{-\infty}^0 
\frac{10^{2i}}{\sqrt{1-10^{2i}}} 
{\rm erfc}\left[\frac{m_l+d-\beta(w-i)-\gamma}{\sqrt{2}\sigma_m}\right] 
10^{(\alpha+1)[\beta(w-i)-v_\star]} \exp (-10^{\beta(w-i)-v_\star}) \, di}
\end{equation}

\item The dual-scatter model:
\begin{equation} \label{eq:cpdf_uTFR_step}
p(\tilde{m} | \tilde{w},d) = 
\frac{\sqrt{2}}{\sqrt{\pi} \sigma_m} 
\frac{\int_{-\infty}^0 
\frac{10^{2i}}{\sqrt{1-10^{2i}}} 
\left( \int_{-\infty}^{\infty} 
\exp \left[-\frac{(w-\tilde{w})^2}{2\sigma_w^2}\right] 
\exp \left[-\frac{(\tilde{m}+d-\beta(w-i)-\gamma)^2}{2\sigma_m^2}\right] 
10^{(\alpha+1)[\beta(w-i)-v_\star]} \exp (-10^{\beta(w-i)-v_\star}) 
dw \right) di}
{\int_{-\infty}^0 \frac{10^{2i}}{\sqrt{1-10^{2i}}} 
\left( \int_{-\infty}^{\infty} 
\exp \left[-\frac{(w-\tilde{w})^2}{2\sigma_w^2}\right] 
{\rm erfc} (\frac{m_l+d-\beta(w-i)-\gamma}{\sqrt{2}\sigma_m})
10^{(\alpha+1)[\beta(w-i)-v_\star]} \exp (-10^{\beta(w-i)-v_\star}) 
dw \right) di}
\end{equation}

\end{itemize}

For the inverse model, simplifying the conditional pdf does not require a step function, but any function that depends only on $m$ [i.e., $S(m,\tilde{w}) = S(m)$]. When this is the case, the selection functions can be taken out of the integrals and cancel out. After the cancellation of $S(m)$, the denominator becomes unity because both $p(\tilde{w}|m,i,d)$ and $p(i)$ are properly normalized pdfs. The result has been given in Eq.\,\ref{eq:cpdf_iTFR_Sm} and is repeated here for completeness:  
\begin{align} \label{eq:cpdf_iTFR_Sm1}
p(\tilde{w}|m,d) &= \frac{S(m) \int_{-\infty}^0 p(\tilde{w}|m,d,i) p(i) \, di}{\int_{-\infty}^0 \left(\int_{-\infty}^{\infty} S(m) p(\tilde{w}|m,d,i) \, d\tilde{w} \right) p(i) \, di} \nonumber \\
&= \int_{-\infty}^0 p(\tilde{w}|m,d,i) p(i) \, di \nonumber \\
&=\frac{\ln 10}{{\sigma_w\sqrt{2\pi}}} \int_{-\infty}^0 \frac{10^{2i}}{\sqrt{1-10^{2i}}} \exp \left[ -\frac{[\tilde{w}-i-(m+d-\gamma)/\beta]^2}{2\sigma_w^2} \right] di 
\end{align}

\newpage
\section{Dual-Scatter Model Starting from the Inverse Model} \label{sec:unify_iTFR}

In this Appendix, I derive the joint PDF of the observables for the latent-inclination dual-scatter model, starting from the inverse model, complementing the derivation in \S\,\ref{sec:uTFR} that began with the forward model.
 
Starting from the joint pdf of $p(\tilde{w},m,d,i)$ in Eq.\,\ref{eq:joint_iTFR2}:
\begin{align} 
p(\tilde{w},m,d,i) &= p(\tilde{w}|m,d,i) p(m|d) p(d) p(i) \nonumber \\
&= p(d) \times \frac{\ln 10 \cdot 10^{2 i }}{\sqrt{1-10^{2 i}}} \frac{1}{{\sigma_w \sqrt {2\pi } }} \exp \left[ -\frac{[\tilde{w}-i-(m+d-\gamma)/\beta]^2 }{2\sigma_w^2 } \right] \phi_\star 10^{(\alpha+1) (m+d-M_\star)} \exp (-10^{m+d-M_\star})
\end{align} 
To account for measurement error in $m$, one needs to multiply an additional term because the joint probability including $\tilde{m}$ is:
\begin{align}
p(\tilde{w},\tilde{m},d,m,i) 
= p(\tilde{m}|m) p(\tilde{w},m,d,i)
\end{align} 
where the new factor is a Gaussian: 
\begin{equation}
p(\tilde{m}|m) = \frac{1}{\sigma_m\sqrt{2\pi}} \exp \left[ -\frac{(\tilde{m}-m)^2}{2\sigma_m^2} \right]
\end{equation}

The joint pdf of the observables $(\tilde{w},\tilde{m},d)$ is then calculated by marginalizing the above joint pdf over both $m$ and $i$:
\begin{align} 
p(\tilde{w},\tilde{m},d) &= \int_{-\infty}^0 \int_{-\infty}^{\infty} p(\tilde{w},\tilde{m},d,m,i)\,dm\,di \nonumber \\
&= \frac{\phi_\star p(d)}{2\pi\sigma_w\sigma_m} \int_{-\infty}^0 \int_{-\infty}^{\infty} 
\exp \left[ -\frac{[\tilde{w}-i-(m+d-\gamma)/\beta]^2}{2\sigma_w^2} \right]
\exp \left[ -\frac{(\tilde{m}-m)^2}{2\sigma_m^2} \right]
10^{(\alpha+1)(m+d-M_\star)} \exp (-10^{m+d-M_\star}) 
p(i)\,dm\,di 
\end{align}

When substituting $m$ for $\beta(w-i)+\gamma-d$ using the idealized TFR, the above equation becomes:
\begin{align} 
&p(\tilde{w},\tilde{m},d) \nonumber \\
&= \frac{\beta \phi_\star p(d)}{2\pi\sigma_w\sigma_m} \int_{-\infty}^0 \int_{-\infty}^{\infty} 
\exp \left[ -\frac{(\tilde{w}-w)^2}{2\sigma_w^2} \right]
\exp \left[ -\frac{[\tilde{m}+d-\beta(w-i)-\gamma]^2}{2\sigma_m^2} \right]
10^{(\alpha+1)[\beta(w-i)-v_\star]} \exp [-10^{\beta(w-i)-v_\star}] 
p(i)\,dw\,di 
\end{align}
which is equivalent to $p(\tilde{m},\tilde{w},d)$ in Eq.\,\ref{eq:joint_uTFR2}. This exercise shows that the data likelihood function of the latent-inclination dual-scatter model remains the same whether one starts from the forward model or the inverse model. This proves that the dual-scatter model is bidirectional (i.e., symmetric), unlike the forward and the inverse models.

\newpage
\section{Numerical Acceleration by Fast Fourier Transform and PyTorch} \label{sec:accelerate}

For the latent-inclination dual-scatter model with step-like selection function, the conditional pdf that constitutes the likelihood function is (Eq.\,\ref{eq:cpdf_uTFR_step} in Appendix\,\ref{sec:step_fun}):
\begin{equation} \label{eq:cpdf_uTFR_step1}
p(\tilde{m} | \tilde{w},d) = 
\frac{\sqrt{2}}{\sqrt{\pi} \sigma_m} 
\frac{\int_{-\infty}^0 
\frac{10^{2i}}{\sqrt{1-10^{2i}}} 
\left( \int_{-\infty}^{\infty} 
\exp \left[-\frac{(w-\tilde{w})^2}{2\sigma_w^2}\right] 
\exp \left[-\frac{(\tilde{m}+d-\beta(w-i)-\gamma)^2}{2\sigma_m^2}\right] 
10^{(\alpha+1)[\beta(w-i)-v_\star]} \exp (-10^{\beta(w-i)-v_\star}) 
dw \right) di}
{\int_{-\infty}^0 \frac{10^{2i}}{\sqrt{1-10^{2i}}} 
\left( \int_{-\infty}^{\infty} 
\exp \left[-\frac{(w-\tilde{w})^2}{2\sigma_w^2}\right] 
{\rm erfc} (\frac{m_l+d-\beta(w-i)-\gamma}{\sqrt{2}\sigma_m})
10^{(\alpha+1)[\beta(w-i)-v_\star]} \exp (-10^{\beta(w-i)-v_\star}) 
dw \right) di}
\end{equation}
Direct integration of the double integrals for each data point is computationally expensive. For instance, on a 2021 Macbook Pro with an M1 Pro chip (8 performance CPU cores and 2 efficiency cores), direct integrations of 1,024 nodes in each integration axis take $\sim$17\,min per MCMC step for 16 walkers and 10,147 data points. This is painfully slow and forces the user to either reduce the number of nodes or to bin the data points, neither of which is desirable. In this Appendix, I describe numerical methods that expedited the computation by thousands of times.

The key is to realize that the integral over $i$ on the numerator of Eq.\,\ref{eq:cpdf_uTFR_step1},
\begin{equation}
\int_{-\infty}^0
\frac{10^{2i}}{\sqrt{1-10^{2i}}}
\exp \left[-\frac{(\tilde{m}+d-\beta(w-i)-\gamma)^2}{2\sigma_m^2}\right] 
10^{(\alpha+1)[\beta(w-i)-v_\star]} \exp (-10^{\beta(w-i)-v_\star})
di
\end{equation}
is a convolution between:
\begin{equation}
f(i) = 
\begin{cases}
    \frac{10^{2i}}{\sqrt{1-10^{2i}}},  & \text{if } i < 0\\
    0,              & \text{otherwise}
\end{cases}
\end{equation}
and
\begin{equation}
g(w-i) = 
\exp \left[-\frac{(\tilde{m}+d-\beta(w-i)-\gamma)^2}{2\sigma_m^2}\right] 
10^{(\alpha+1)[\beta(w-i)-v_\star]} \exp (-10^{\beta(w-i)-v_\star})
\end{equation}

The result is a new function of $w$: 
\begin{equation}
\{f * g\}(w) = \int_{-\infty}^{\infty} f(i) g(w-i) di
\end{equation}
which can be calculated with Fourier transform given the convolution theorem: 
\begin{equation}
\{f * g\}(w) = \mathcal{F}^{-1} [\mathcal{F}\{f\}(w) \cdot \mathcal{F}\{g\}(w)]
\end{equation}
Since the computation of the Fast Fourier Transform (FFT) of $f$ and $g$ and the inverse FFT of their product has a total complexity of $\mathcal{O}(N\log N)$ and direct computation of the convolution integral has a complexity of $\mathcal{O}(N^2)$, substantial improvement in computing efficiency can be achieved with FFT-based convolution when the number of nodes in each integration axis ($N$) is a large number. 

The result of the convolution can then be supplied to the integral over $w$ to complete the calculation of the numerator:
\begin{equation}
\int_{-\infty}^{\infty} \{f*g\}(w) \exp \left[-\frac{(w-\tilde{w})^2}{2\sigma_w^2}\right] dw
\end{equation}
Notice that even though this second integral is also a convolution, direct evaluation is faster because the convolution result is needed only at a specific value of $\tilde{w}$ instead of an array of $\tilde{w}$. In such cases, FFT methods have no computational advantage over direct evaluation. 

The same method can be applied to compute the denominator, simply replace the function of $g$ as:
\begin{equation}
g(w-i) = {\rm erfc} \left(\frac{m_l+d-\beta(w-i)-\gamma}{\sqrt{2}\sigma_m}\right)
10^{(\alpha+1)[\beta(w-i)-v_\star]} \exp (-10^{\beta(w-i)-v_\star})
\end{equation}

On the same 2021 Macbook Pro with the same 1,024 nodes in each integral, the FFT-accelerated dual-scatter model takes 1.65\,s per MCMC step for 16 walkers and 10,147 data points, which is $\sim$600$\times$ faster than direct integration. The FFT-accelerated dual-scatter model is only 1.9$\times$ and 7.6$\times$ slower than the much simpler forward and inverse models, respectively.

An additional $\sim$4$\times$ acceleration is achieved by vectorizing the function that computes the conditional pdf in Eq.\,\ref{eq:cpdf_uTFR_step1} and leveraging the integrated 16-core Graphics Processing Unit (GPU) with PyTorch's Metal Performance Shaders (MPS) backend \citep{Paszke19}. The GPU- and FFT-powered dual-scatter model takes 0.44\,s per MCMC step, which is even 2$\times$ faster than the simpler forward model on the CPU. The MCMC chains sufficiently converged after $\sim$6,000 steps, so the total computing time is $\sim$0.75\,hr. The PyTorch version of the code is made available along with the NumPy version. 

\end{document}